\documentclass{stsci_report}

\usepackage{graphicx}
\usepackage{siunitx}
\usepackage{todonotes}
\usepackage{pdfpages}
\usepackage{hyperref}
\usepackage{amssymb}
\usepackage{caption}
\usepackage{wrapfig}
\usepackage{xcolor}
\usepackage{tcolorbox}
\usepackage{setspace}
\usepackage{listings}
\usepackage{indentfirst}
\usepackage{amsmath}
\usepackage{amstext}
\usepackage{latexsym}
\usepackage{longtable}
\usepackage{times}
\usepackage{xspace}
\usepackage{multirow}
\usepackage{array}
\usepackage{aas_macros}
\usepackage{natbib}
\usepackage{xcolor}
\defcitealias{2024wfc..rept....6M}{MM24}

\copyrighttext{Copyright\copyright\ \the\year\ The Association of Universities for Research in Astronomy, Inc. All Rights Reserved.}

\presubtitle{Instrument Science Report WFC3 2024-13}
\title{New Time-Dependent WFC3/IR\\ Inverse Sensitivities}
\author{Annalisa Calamida, Mariarosa Marinelli, Varun Bajaj, Aidan Pidgeon, Jennifer Mack}
\date{December 18, 2024}

\begin{document}

\maketitle
\abstract{
We present new time-dependent inverse sensitivities for the WFC3/IR channel. These were calculated using the sensitivity change slopes measured by \citet{2024wfc..rept....6M} and photometry of five CALSPEC standards (the white dwarfs GRW+70~5824, GD~153, GD~71, G191B2B, and the G-type star P330E) collected from 2009 to 2023.
The new inverse sensitivities account for losses of 1-2\% over 15 years, depending on wavelength, and provide an internal photometric precision better than 0.5\% for all wide--, medium--, and narrow-band filters. 
An updated version of \texttt{calwf3} (v3.7.2) has been developed for use with a new time-dependent image photometry table (IMPHTTAB) and will be used to update the image header photometric keywords following MAST reprocessing, expected in late-2024. Alternatively, the new inverse sensitivities may be computed by the user for a specific observation date by running \texttt{stsynphot}.}


\newpage
\tableofcontents

\newpage

\section{Introduction}
WFC3/IR inverse sensitivities enable the conversion of measured fluxes from count rates to physical units; these are saved in the image headers as the PHOTFLAM keyword, and were last updated in 2020 for inclusion in the WFC3 calibration pipeline, \texttt{calwf3} version 3.5.2 (see \citealt{bajaj2020} and \citealt{calamida2022}). 
These were calculated by using updated CALSPEC models\footnote{See the CALSPEC web page for access to the models and more details: \href{https://www.stsci.edu/hst/instrumentation/reference-data-for-calibration-and-tools/astronomical-catalogs/calspec}{https://www.stsci.edu/hst/instrumentation/} \\ \href{https://www.stsci.edu/hst/instrumentation/reference-data-for-calibration-and-tools/astronomical-catalogs/calspec}{reference-data-for-calibration-and-tools/astronomical-catalogs/calspec}} for the three \emph{HST} primary spectrophotometric standard white dwarfs (WDs), GD153, GD71, and G191B2B \citep{bohlin2020} 
and nine years of photometric measurements of these WDs plus the secondary WD standard GRW+70~5824 (hereinafter GRW70), and the G-type standard P330E. 

These inverse sensitivities do not include a time dependence, i.e. the zeropoint (ZP) is calculated for a reference epoch only, a Modified Julian Date (MJD) of 55008 (June 26, 2009), based on the mean over all observation dates.
To evaluate the sensitivity change of the IR detector, the WFC3 team used 
photometry of fields centered on uncrowded regions of Galactic globular clusters, spatial scan observations of an open cluster, and grism spectra of four CALSPEC standard WDs. 

As reported in \citet[hereafter MM24]{2024wfc..rept....6M}, the sensitivity change of the WFC3/IR detector amounts to a total decline of $\approx 1-2 \%$ since installation in 2009. 
Sensitivity losses were found to be wavelength-dependent, with the greatest losses in the bluest filters. 
The sensitivity loss rates were determined as a function of filter pivot wavelength and are reproduced from \citetalias{2024wfc..rept....6M} in Table~\ref{table:1}. 
These previously reported slopes allowed users to manually correct their photometry for time-dependent losses, ahead of the delivery of the updated reference files and an accompanying release of \texttt{calwf3} compatible with those files.

This report discusses the delivery and testing of new time-dependent inverse sensitivities, as well as the release of a new image photometry lookup table (the IMPHTTAB reference file) to be used with the updated WFC3 calibration pipeline, \texttt{calwf3} version 3.7.2, and new time-dependent filter throughputs to be used for simulations with \texttt{stsynphot} \footnote{ \href{https://stsynphot.readthedocs.io/en/latest/stsynphot/obsmode.html}{https://stsynphot.readthedocs.io/en/latest/} }.

\clearpage
\section{New filter throughputs} \label{sec:filt_corr}


To calculate new inverse sensitivities for the WFC3/IR detector, we used staring mode photometry of five CALSPEC standard stars: GD153, GD71, G191B2B, GRW70, and P330E. 
These stars are regularly observed as part of WFC3's calibration monitoring programs and the data used for this analysis span the time interval of 2009 -- 2023. For a list of the program IDs and more details on the number of observations collected for each target in different filters see Appendix A in \citetalias{2024wfc..rept....6M}.

In Section~\ref{sec:corr_phot}, we discuss how the observed count rates of the standard stars for all 15 filters were corrected for time-dependent sensitivity losses according to the slopes reported in \citetalias{2024wfc..rept....6M}. 
In Section~\ref{sec:cr_ratios}, we describe how the corresponding synthetic observations were generated and how the observed-to-synthetic count rate ratios for the standard stars were calculated.
From the corrected count rate ratios, we then derived multiplicative scalar factors to be applied to the filter curves, as detailed in Section~\ref{sec:filt_curves}. 

\begin{minipage}[!b]{0.9\linewidth} 
\vspace{2ex}
    \centering 
    \def\arraystretch{1.2} 
    \begin{tabular}[!b]{|c|c|c|c|}
    \hline
\textbf{Filter} & \textbf{$\lambda_{p}$} \textbf{(nm)} & \textbf{$m_{corr}$} \textbf{(\%/year)} & \textbf{$\Delta \phi_{t_{i}-t_{0}}$ (\% in 15 years)} \\ 
\hline
 F098M & 986.4  & $0.120 \pm 0.003$ & $\approx 1.68$ \\ \cline{1-2}
 F105W & 1055.5 & &  \\  \cline{1-2}
 F110W & 1153.4 & &  \\  \hline
F125W & 1248.6 & $0.075 \pm 0.006$ & $\approx 1.05$ \\ \cline{1-2}
F126N & 1258.5 & & \\ \cline{1-2}
F127M & 1274.0 & & \\ \cline{1-2}
F128N & 1283.2 & & \\ \cline{1-2}
F130N & 1300.6 & & \\ \cline{1-2}
F132N & 1318.8 & & \\ \hline

F139M  & 1383.8  & $0.060 \pm 0.005$ &  $\approx 0.84$ \\ \cline{1-2}
F140W  & 1392.3 & &  \\ \cline{1-2}
F153M  & 1532.2 & &  \\ \cline{1-2}
F160W & 1536.9 & &  \\ \cline{1-2}
F164N & 1640.4 & &  \\ \cline{1-2}
F167N & 1664.2 & &  \\ \hline

    \end{tabular}  
\captionsetup{type=table}
\captionof{table}{All 15 WFC3/IR filters, their pivot wavelengths ($\lambda_{p}$, in nm), and the corresponding time-dependent sensitivity correction factors ($m_{corr}$, in $\% /$ year).  
The total decrease in the measured flux since installation 15 years ago, attributable to sensitivity losses, is given in the last column $\Delta \phi_{t_{i}-t_{0}}$. Table reproduced from \citet{2024wfc..rept....6M}. \label{table:1}}
\vspace{2ex}
\end{minipage}

\subsection{Correcting the standard star photometry} \label{sec:corr_phot}

The photometry of the five CALSPEC standard stars was corrected for sensitivity losses according to the slopes listed in Table~\ref{table:2}.
After correcting the observed photometry, we calculated the mean count rate for each standard star at the reference epoch (MJD $= 55008$) for all 15 filters.

\subsection{Comparison to synthetic photometry} \label{sec:cr_ratios}

We calculated synthetic count rates at the reference epoch for each CALSPEC standard star by using the software package \texttt{stsynphot} \citep{2020ascl.soft10003S}.
For these simulations, we adopted the most recent CALSPEC spectral energy distributions (SEDs; \citealt{bohlin2020}), the latest set of WFC3/IR throughput curves, including the 2009 in-flight quantum efficiency correction file, the 2020 filter curves, and the 2009 aperture correction file. 
File names used for the synthetic simulations and their descriptions are listed in Table~\ref{table:2}. 

\begin{figure}[!b]
\centering
\includegraphics[width=0.6\linewidth, angle=90]{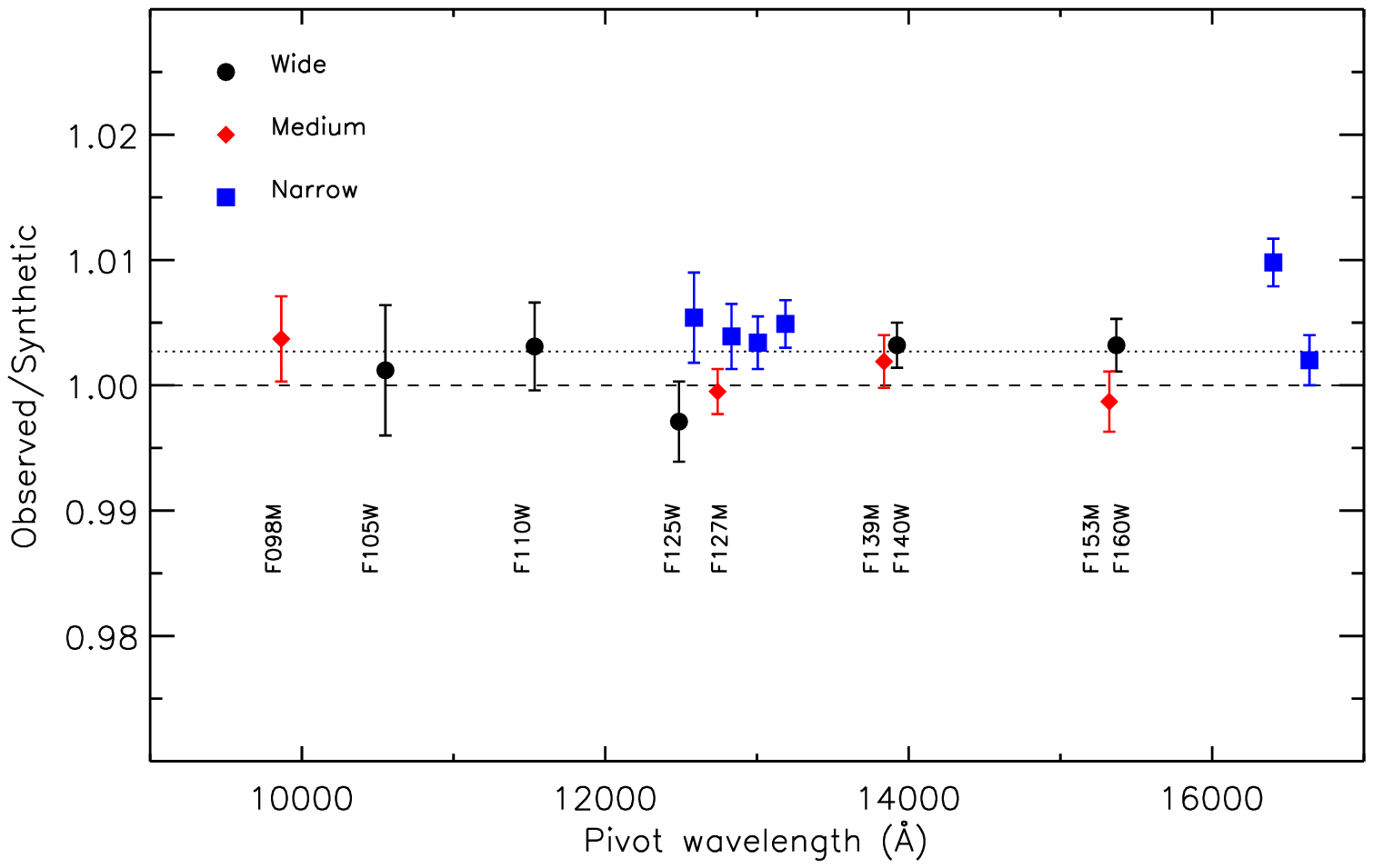} 
\captionsetup{type=table, width=.9\linewidth}
\captionof{figure}{Ratio of the observed to synthetic count rates, calculated as the weighted mean of the five CALSPEC standard stars in the 15 WFC3/IR filters: wide (black circles), medium (red diamonds), and narrow (blue squares). Error bars are displayed and only wide- and medium-band filters are labeled. The dashed black line indicates a unity ratio, and the dotted line shows the mean ratio over all filters. \label{fig:cor}
}
\end{figure}

We then calculated the mean ratio of the observed to synthetic count rates for the five standard stars for each filter, after weighting for photometric errors. 
These are plotted as a function of the filter pivot wavelength in Fig.~\ref{fig:cor}. 
The ratio for most filters (except F125W, F127M and F153M, marked in the figure) is slightly larger than 1.0, indicating that the filter throughputs were underestimated in the latest calibration \citep{bajaj2020}. 
This result is expected, since the observed to synthetic count rate ratios were calculated at that time without correcting the standard star photometry for sensitivity losses. 
On the other hand, the mean ratio over all 15 filters is 1.003$\pm$0.003 (dotted line in the figure), demonstrating the high level of accuracy of the previous calibration. The incorporation of a time-dependent correction improves this accuracy further, as will be shown in this report.

\begin{minipage}[!b]{0.9\linewidth} 
\vspace{3ex}
\centering
\def\arraystretch{1.15}
\begin{tabular}{|c|l|c|}
\hline
\textbf{Purpose} & \multicolumn{1}{c|}{\textbf{Component}} & \textbf{Description} \\
\hline
& \texttt{wfc3\_ir\_cor\_004\_syn.fits} & In-flight correction (2009) \\ \cline{2-3}
& \texttt{wfc3\_ir\_aper\_002\_syn.fits} & Aperture correction (2009) \\ \cline{2-3}
Simulations & \texttt{wfc3\_ir\_FXXXX\_007\_syn.fits} & Current filter curves (2020) \\ \cline{2-3}
to derive the & \texttt{gd153\_stiswfcnic\_004.fits}  &   \\ \cline{2-2}
new filter & \texttt{gd71\_stiswfcnic\_004.fits}  &  \\ \cline{2-2}
curves & \texttt{gd191b2b\_stiswfcnic\_004.fits}       &  CALSPEC SED \\ \cline{2-2}
& \texttt{grw\_70d5824\_stiswfcnic\_003.fits}   &   \\ \cline{2-2}
& \texttt{p330e\_stiswfcnic\_007.fits}             &   \\ \cline{2-3}
\hline
&  \texttt{wfc3\_ir\_cor\_004\_syn.fits} & In-flight correction (2009) \\ \cline{2-3}
&  \texttt{wfc3\_ir\_aper\_002\_syn.fits} & Aperture correction (2009) \\ \cline{2-3}
Simulations  &  \texttt{wfc3\_ir\_FXXXX\_008\_syn.fits} & New filter curves (2024) \\ \cline{2-3}
to derive the  &  \texttt{gd153\_stiswfcnic\_004.fits}  &    \\  \cline{2-2}
final synthetic  &  \texttt{gd71\_stiswfcnic\_004.fits}  &   \\  \cline{2-2}
count rates  &  \texttt{gd191b2b\_stiswfcnic\_004.fits}  &  CALSPEC SED \\  \cline{2-2}
&  \texttt{grw\_70d5824\_stiswfcnic\_003.fits}  &    \\  \cline{2-2}
&  \texttt{p330e\_stiswfcnic\_007.fits}  &    \\  \cline{2-2}
\hline
\end{tabular}
\captionsetup{type=table}
\captionof{table}{Files used in the synthetic simulations performed with \texttt{stsynphot}. These are available to download from the HST Calibration Reference Data System (CRDS) at \href{https://hst-crds.stsci.edu/}{https://hst-crds.stsci.edu/}} \label{table:2}
\end{minipage}
\vspace{1ex}

\begin{figure}[!t]
\centering
\includegraphics[width=0.6\linewidth, angle=90]{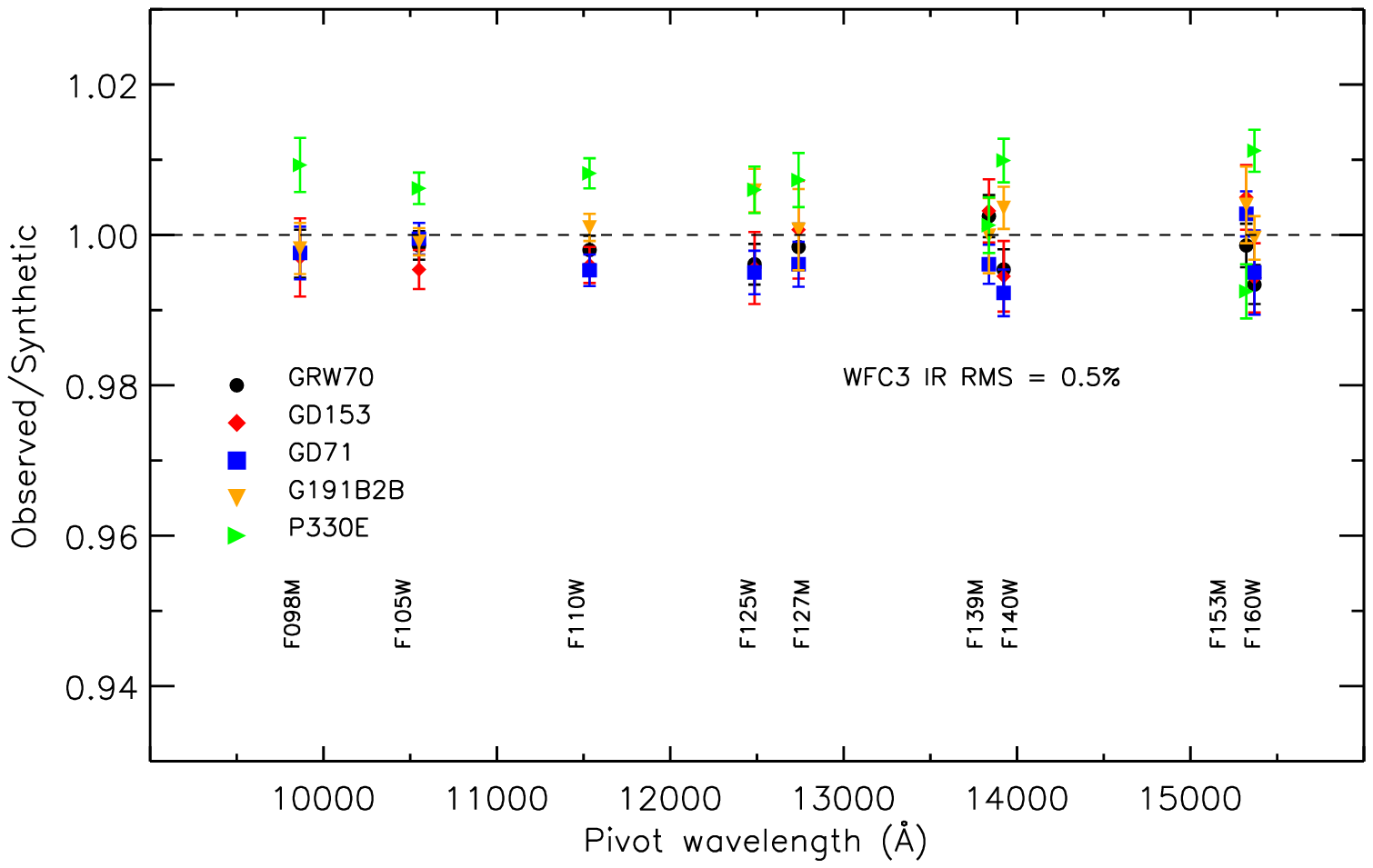} 
\captionsetup{type=table, width=.9\linewidth}
\captionof{figure}{Ratio of the observed to synthetic count rates in five wide-band and four medium-band WFC3/IR filters as a function of pivot wavelength for the five CALSPEC standards used in the calibration. The dashed black line indicates a unity ratio. Error bars are displayed. \label{fig:ratio}}
\end{figure}

\subsection{Filter curves} \label{sec:filt_curves}

The weighted mean of the observed to synthetic count rate ratio for the five CALSPEC standard stars for each filter was used to derive a multiplicative scalar correction to be applied to the filter curves.
The updated filter curves (\texttt{wfc3\_ir\_FXXXX\_008\_syn.fits}) were used to calculate the synthetic count rates for each standard star and filter at the reference epoch of MJD = 550008. 
Time-dependent filter curves (\texttt{wfc3\_ir\_FXXXX\_mjd\_008\_syn.fits}) were also created in order to provide the expected count rates of the standard stars at different epochs. 
Note that when using \texttt{stsynphot} 
to derive count rates for a given target at different epochs, the code interpolates between pairs of consecutive MJD values included in the filter curve tables. 
If the requested epoch is outside the current lifetime of WFC3/IR, the values will be extrapolated into the future (or into the past).
The extrapolation to MJD values before the reference epoch, i.e. before WFC3 was installed, is not reliable and should not be used in simulations.

Fig.~\ref{fig:ratio} shows the observed to synthetic count rate ratios for the five standard stars obtained using the new filter curves for all wide- and medium-band filters.
The ratios cluster around $1.0$, as expected, with a RMS of $0.5\%$. 
There is a clear improvement in comparison to the previous calibration released in 2020, where the standard star photometry was not corrected for any time-dependence. 
In that case, the observed to synthetic count rate ratios show a larger dispersion, with a total RMS of 0.6\% (see Fig.~15 of \citealt{calamida2022}).

It is worth noting that the ratios obtained for P330E are systematically higher compared to those obtained for the WDs for all filters except F153M (as they were in the 2020 calibration).
The SED of the CALSPEC standard P330E is based on the BOSZ models, recently updated by \citet{meszaros2024}, combined with STIS, WFC3, and NICMOS observations.
In the WFC3/IR wavelength range, the SED for P330E is solely defined by its WFC3 grism observations, including 8 collected with G102 and 13 with G141. 
Since flat field errors for WFC3/IR amount on average to $\approx$ 1\% \citep{mack2021}, there could be a small systematic error due to the placement of the P330E grism observations on a brighter spot of the flat field, for instance. 
On the other hand, the four standard WDs have about one order of magnitude more grism spectra collected at different locations on the detector, compared to P330E, and should be less affected by flat field uncertainties (see also \citealt{som2024}). 
This issue is under investigation by the WFC3 team. 
Note that if P330E is eliminated, the mean observed over synthetic count rate ratio changes by less than 0.1\%, a much smaller shift compared to the relative precision , $\approx$ 0.5 \%, of the provided inverse sensitivities.

\clearpage
\section{Calculating the inverse sensitivities\label{sec:sensi}}

We now have updated synthetic count rates and time-dependent corrected observed count rates for the five standard stars in 15 WFC3/IR filters.
In order to derive new inverse sensitivities, we followed the method of \citet{calamida2022}.
We briefly report here some basic equations to better understand how the inverse sensitivities are derived.
The instrument count rate, $N_e$, in photoelectrons per second, is defined as:

\begin{equation}
N_{e} = \frac{A}{hc}\int F_{\lambda }\cdot \lambda \cdot R\cdot d\lambda 
\end{equation}

\medskip

where $A$ is the telescope collecting area, $h$ is the Planck constant, $c$ is the speed of light, $R$ is the system throughput, and $F_{\lambda}$ is the source spectral flux density in erg $\cdot$ s$^{-1}$ $\cdot$ cm$^{-2}$ $\cdot$ \AA$^{-1}$.

The sensitivity $S$, which is defined as the flux that gives 1 count/s, can now be written as:

\begin{equation}
S = \frac{hc}{A\cdot \int \lambda \cdot R\cdot d\lambda }
\end{equation}

\medskip

and has units of erg $\cdot$ cm$^{-2}$ $\cdot$ \AA$^{-1} /$ e$^-$. The flux is thus:

\begin{equation}
F = S\cdot N_{e} = S\cdot C
\end{equation}

\medskip

where $C$ is the source count rate in e$^{-}$ $\cdot$ s$^{-1}$. We can derive the inverse sensitivity, $S'$, by inverting the last relation:

\begin{equation}
S' = 1/S = C / F
\end{equation}

\medskip

where $C$ is available from the observations of the standard stars and $F$ is provided by the \texttt{stsynphot} simulations performed with the updated filter curves\footnote{Updated filter curve filenames follow the convention \texttt{wfc3\_ir\_FXXXX\_008\_syn.fits}, as seen in Table \ref{table:2}.}. 

New inverse sensitivities for the reference epoch (MJD $= 55008$) in three photometric systems, namely AB, Vega, and ST, are listed in Table \ref{table:3} and can be found on the 
\\
\href{https://www.stsci.edu/hst/instrumentation/wfc3/data-analysis/photometric-calibration/ir-photometric-calibration}{WFC3/IR Photometric Calibration web page}.\footnote{IR Photometric Calibration: \href{https://www.stsci.edu/hst/instrumentation/wfc3/data-analysis/photometric-calibration/ir-photometric-calibration}{https://www.stsci.edu/hst/instrumentation/wfc3/data-analysis/photometric-calibration/} \href{https://www.stsci.edu/hst/instrumentation/wfc3/data-analysis/photometric-calibration/ir-photometric-calibration}{ir-photometric-calibration}} 
These inverse sensitivities were calculated at an effectively infinite aperture radius, thus encompassing all the flux, and were derived after correcting the photometry from a 3-pixel aperture radius to infinity using the current Encircled Energy (EE) corrections.\footnote{IR Encircled Energy: \href{https://www.stsci.edu/hst/instrumentation/wfc3/data-analysis/photometric-calibration/ir-encircled-energy}{https://www.stsci.edu/hst/instrumentation/wfc3/data-analysis/photometric-calibration/ir-} \href{https://www.stsci.edu/hst/instrumentation/wfc3/data-analysis/photometric-calibration/ir-encircled-energy}{encircled-energy}}. 
Inverse sensitivities as a function of observation date are populated in the \texttt{PHOTFLAM} and \texttt{PHOTNU} header keywords of WFC3/IR image FITS files, as of \texttt{calwf3} v3.7.2. 
Alternately, they can be computed ``on-the-fly’' for any observing epoch by using \texttt{stsynphot} and the new set of filter curves (see Table \ref{table:2}). For guidance, users may wish to consult the Jupyter notebook tutorial \href{https://spacetelescope.github.io/hst_notebooks/notebooks/WFC3/zeropoints/zeropoints.html}{Calculating WFC3 Zeropoints with stsynphot.}\footnote{Zeropoints notebook: \href{https://spacetelescope.github.io/hst_notebooks/notebooks/WFC3/zeropoints/zeropoints.html}{https://spacetelescope.github.io/hst\_notebooks/notebooks/WFC3/zeropoints/zeropoints.html}}



We then created a new IMPHTTAB reference file for use with \texttt{calwf3} by adopting the calculated inverse sensitivities at different epochs, and the updated pivot wavelengths and bandwidths for all 15 WFC3/IR imaging filters.

\medskip

\begin{minipage}[!b]{1\linewidth} 
\vspace{2ex}
\centering
\def\arraystretch{1.1}
\footnotesize
\begin{tabular}{lcccccccc}
\hline
\hline
Filter & PHOTPLAM & PHOTBW  & ZP$_{AB}$  &  ZP$_{Vega}$ &  ZP$_{ST}$  &  ZP$_{ERR}$ & PHOTFLAM & PHOTFLAM$_{\text{ERR}}$ \\
    &  (\AA)  & (\AA)   & (Mag) & (Mag) & (Mag) & (Mag)  & (erg cm$^{-2}$ & (erg cm$^{-2}$ \\
    &    &    &   &   &   &  & \AA$^{-1}$ e$^{-1}$) & \AA$^{-1}$ e$^{-1}$) \\ \hline
\hline
F098M &  9864.72 &  500.85 & 25.6684 & 25.0923 & 26.9468 & 0.0055 & 6.0435e-20 & 8.4188e-23  \\
F105W & 10551.05 &  845.62 & 26.2656 & 25.6043 & 27.6900 & 0.0039 & 3.0479e-20 & 2.3774e-23  \\
F110W & 11534.46 & 1428.48 & 26.8223 & 26.0456 & 28.4402 & 0.0048 & 1.5273e-20 & 1.1517e-23  \\
F125W & 12486.06 &  866.28 & 26.2288 & 25.3085 & 28.0189 & 0.0061 & 2.2512e-20 & 2.6317e-23  \\
F126N & 12584.89 &  339.31 & 22.8584 & 21.9176 & 24.6656 & 0.0086 & 4.9400e-19 & 7.8474e-22  \\
F127M & 12740.29 &  249.56 & 24.6266 & 23.6452 & 26.4604 & 0.0045 & 9.4582e-20 & 1.3457e-22  \\
F128N & 12831.84 &  357.44 & 22.9593 & 21.9014 & 24.8087 & 0.0082 & 4.3298e-19 & 6.8757e-22  \\
F130N & 13005.68 &  274.24 & 22.9821 & 21.9856 & 24.8607 & 0.0053 & 4.1279e-19 & 6.3957e-22  \\
F132N & 13187.71 &  319.08 & 22.9359 & 21.9178 & 24.8447 & 0.0060 & 4.1892e-19 & 6.6561e-22  \\
F139M & 13837.62 &  278.03 & 24.4660 & 23.3660 & 26.4793 & 0.0031 & 9.2958e-20 & 1.1698e-22  \\
F140W & 13922.91 & 1132.38 & 26.4540 & 25.3567 & 28.4806 & 0.0079 & 1.4713e-20 & 1.7008e-23  \\
F153M & 15322.05 &  378.95 & 24.4455 & 23.1697 & 26.6800 & 0.0057 & 7.7263e-20 & 1.0218e-22  \\
F160W & 15369.18 &  826.25 & 25.9409 & 24.6669 & 28.1821 & 0.0081 & 1.9370e-20 & 2.3646e-23  \\
F164N & 16403.51 &  700.06 & 22.8997 & 21.4913 & 25.2824 & 0.0053 & 2.7993e-19 & 4.1730e-22  \\
F167N & 16641.60 &  645.24 & 22.9420 & 21.5559 & 25.3560 & 0.0106 & 2.6157e-19 & 3.5870e-22  \\
\hline
\hline

\end{tabular}
\captionsetup{type=table}
\captionof{table}{New inverse sensitivity values (\texttt{PHOTFLAM}) and zeropoints (ZPs) in different photometric systems for the 15 WFC3/IR filters calculated at the reference epoch MJD $= 55008$  (June 26, 2009). \texttt{PHOTFLAM} errors are also given, along with the filter pivot wavelengths (\texttt{PHOTPLAM}) and RMS bandwidths (\texttt{PHOTBW}). \label{table:3}}
\end{minipage}
\vspace{1ex}

\subsection{Comparison with previous inverse sensitivity values\label{sec:comp}}
 
We compared the new inverse sensitivities and zeropoints (ZPs) with those from the previous (2020) photometric calibration. 
These latter ZPs were calculated by using 10 years of available data (2009 -- 2019) for the five CALSPEC standard stars, and measurements were simply averaged without correcting for any changes with time. 
In this work, the inverse sensitivities were derived using approximately 14 years of available data (2009 -- 2023), and the photometric measurements of the standard stars are now normalized to a single reference epoch.

\begin{figure}[!b]
\centering
\includegraphics[height=0.5\textheight, angle=90]{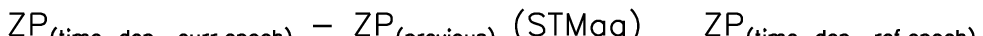} 
\vspace{-0.3cm}
\captionsetup{type=table, width=.9\linewidth}
\captionof{figure}{Top: Comparison between the new time-dependent and previous ZPs for WFC3/IR in the ST photometric system for the wide- (black circles) and medium-band (red diamonds) filters, calculated for the reference epoch (June 2009), plotted against pivot wavelength. A dotted line indicates a ZP difference of zero, and the dashed/dotted line 
shows the mean ZP difference over all filters.
Error bars are displayed.
Bottom: Same comparison but the new time-dependent ZPs were calculated for the current epoch (July 2024).
\label{fig:zp_comp}}
\end{figure}

The top panel of Fig.~\ref{fig:zp_comp} shows the difference between the new time-dependent ZPs calculated at the reference epoch and the ZPs from the previous calibration, computed in the ST photometric system, as a function of pivot wavelength for the wide- and medium-band filters. 
The ZPs differ on average by 0.16\%, with the new values being slightly fainter in 6 out of 9 filters.
The difference between the two sets of ZPs is due to two main factors: the standard star photometry being corrected for losses in sensitivity (which brightens the ZPs) and changes in the CALSPEC standard SEDs, which can brighten or dim the ZPs according to the star and the wavelength regime. 
To demonstrate this, the bottom panel of Fig.~\ref{fig:zp_comp} shows the difference between the new time-dependent ZPs calculated at the current epoch, July 2024, and the previous ZPs; in this case the new ZPs are systematically brighter than the previous ones. 
The difference is larger, $\approx$ 2\% (or 0.02 mag), at bluer wavelengths where the sensitivity loss (and the applied time-dependent correction) is larger, while the difference is smaller ($\lesssim$ 1\% or 0.01 mag) at longer wavelengths where those quantities are smaller (see Table~\ref{table:2}).

Major changes to the standard star SEDs since the photometric calibration of 2020 \citep{bohlin2020} are due to a new CTE correction for the STIS data \citep{bohlin2022}, updated BOSZ models \citep{meszaros2024}, and a new extinction law \citep{gordon2023}. 

Summarizing, the new inverse sensitivities and ZPs for WFC3/IR at the reference epoch do not significantly change compared to the previous (2020) values.
However, a time dependence of the inverse sensitivities was added to the newest version of the WFC3 calibration pipeline (\texttt{calwf3} v3.7.2), which now populates the \texttt{PHOTFLAM} and \texttt{PHOTFNU} header keywords of WFC3/IR image FITS files with the appropriate inverse sensitivity values for the epoch of the observation. 
Users should use these updated header values to flux calibrate their photometry.

\clearpage
\section{Testing the new time-dependent inverse sensitivities} \label{testing}

We performed a first validation of the new time-dependent IMPHTTAB and of the updated calibration pipeline (\texttt{calwf3} v3.7.2) using
observations of the standard GD153 collected in all WFC3/IR filters in the time range 2009 -- 2023.  We downloaded the uncalibrated (``RAW'') images from the Mikulski Archive for Space Telescopes (MAST) and used the new IMPHTTAB file to process them with \texttt{calwf3} v3.7.2, thus generating flat-fielded calibrated (``FLT'') images. 

\begin{figure}[!b]
\centering
\includegraphics[width=0.59\linewidth, angle=90]{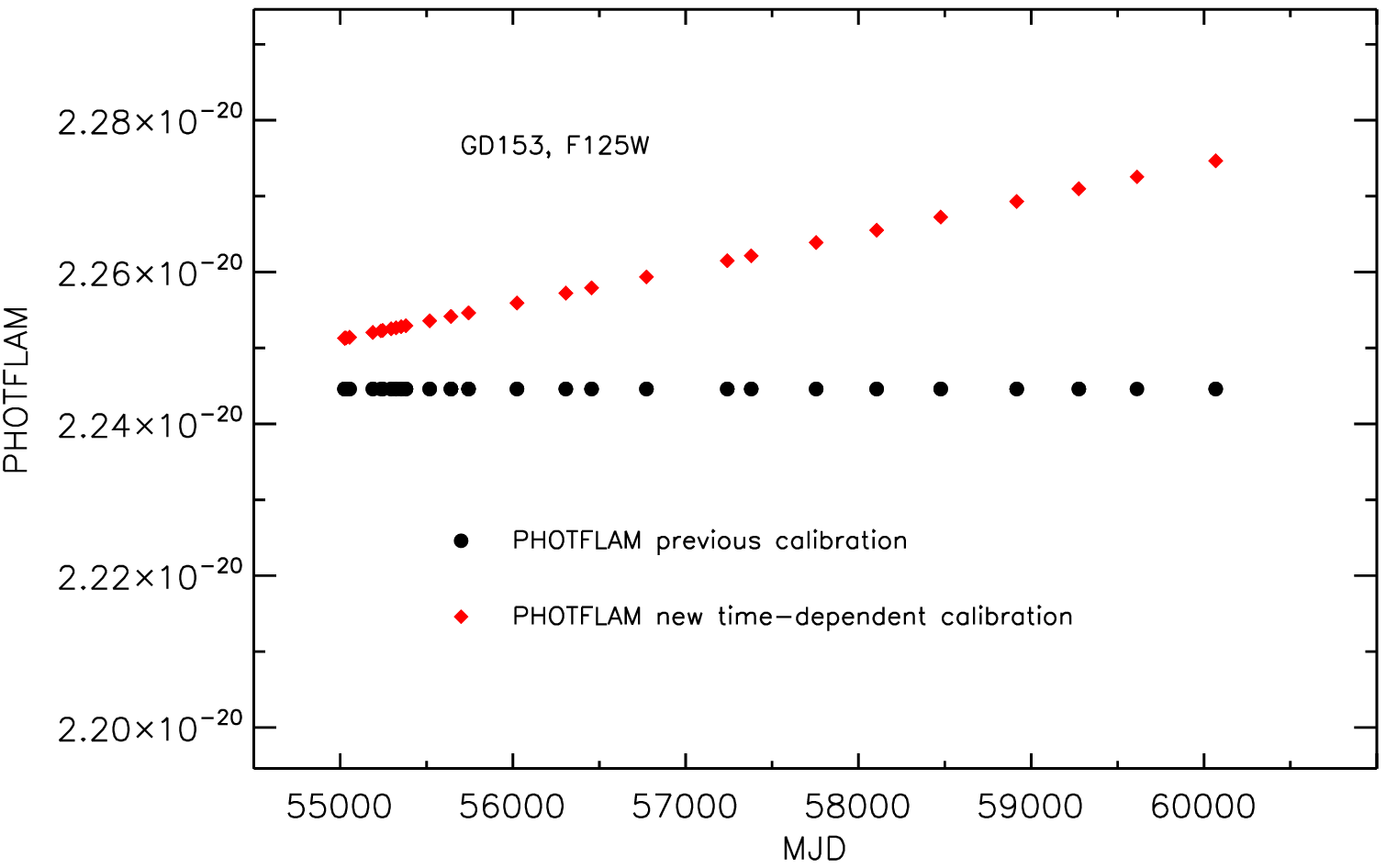}
\captionsetup{type=table, width=.9\linewidth}
\captionof{figure}{New time-dependent \texttt{PHOTFLAM} values extracted from the header of the FLT images of the standard WD GD153 observed over multiple epochs in the F125W filter (filled red diamonds) compared to the constant \texttt{PHOTFLAM} value from the previous photometric calibration (filled black circles). \label{fig:comp}}
\end{figure}

Fig.~\ref{fig:comp} shows the \texttt{PHOTFLAM} values extracted from the header of the F125W images of GD153 as a function of observing epoch (red diamonds) compared to the constant \texttt{PHOTFLAM} values of the previous calibration (black circles). 
This plot clearly demonstrates that the new version of \texttt{calwf3} executes correctly, populating the image headers with time-dependent \texttt{PHOTFLAM} values. 
It is worth noting that \texttt{PHOTFLAM} inverse sensitivity values increase as the detector sensitivity decreases with time, as seen in Fig.~\ref{fig:comp}.

\begin{figure}[!t]
\centering
\includegraphics[width=0.59\linewidth,  angle=90]{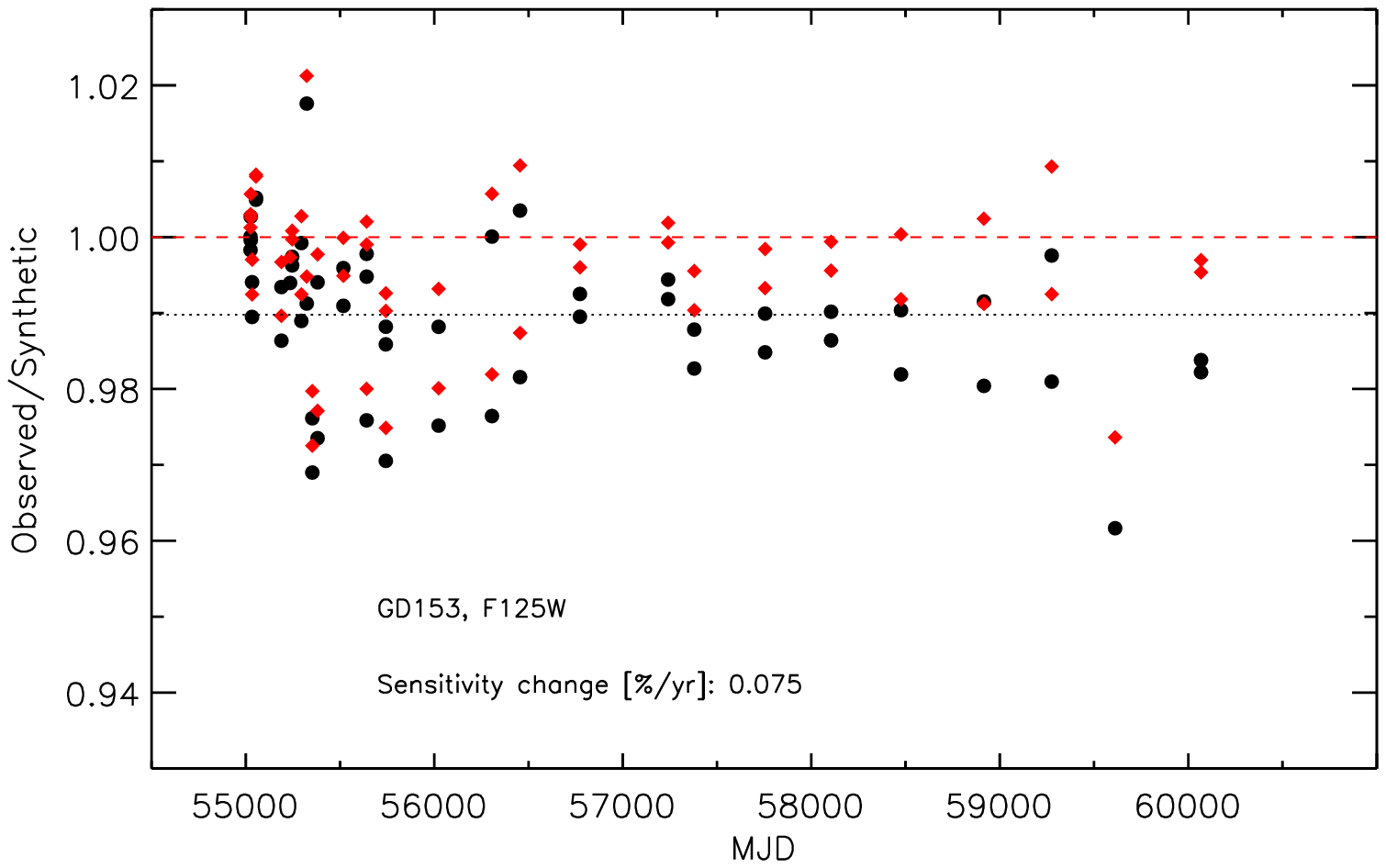}
\captionsetup{type=table, width=.9\linewidth}
\captionof{figure}{Observed relative to synthetic fluxes in the F125W filter for the standard WD GD153 as a function of epoch after calibrating the observations with the time-dependent \texttt{PHOTFLAM} values (filled red diamonds) or with the constant \texttt{PHOTFLAM} value from the previous calibration (filled black circles). The dashed red line and the dotted black line indicate the unity value and the mean one, respectively. \label{fig:flam_gd153}}
\end{figure}

As a further test, we used these \texttt{PHOTFLAM} values to compare the GD153 fluxes measured at different epochs in the F125W filter. 
Fig.~\ref{fig:flam_gd153} shows the observed to synthetic flux ratio obtained by multiplying the observed count rates with the constant \texttt{PHOTFLAM} value from the previous calibration (filled black circles). The same ratio is calculated after correcting the count rates for sensitivity changes (0.075\%/yr) by using the new time-dependent \texttt{PHOTFLAM} values (filled red diamonds). 
The mean of the old (black circles) and the new (red diamonds) observed to synthetic flux ratios is 0.990 and 0.996, respectively, in good agreement with the expected sensitivity change in the time period covered by the observations, $\approx$ 13.5 years, i.e. $\approx$ 1\%. 
It is worth noting that the dispersion of the measurements decreases from 0.010 to 0.009, demonstrating that correcting for sensitivity changes improves the precision of the photometry.

We performed the same test for all the WFC3/IR filters obtaining very similar results.

\subsection{Photometric observations}
In order to verify the reliability and the precision of the new WFC3/IR time-dependent photometric calibration, we used data collected with the F127M and F153M filters for a region towards the Galactic center ($Sgr A^{*}$) over a time interval of $\approx$ 14-years (GO programs 11671, 12318, 12667, 13049, 13316, 13403, 15199, 15498, 15667, 15894, 16004, 16681, and 16990).

Point-spread function (PSF) photometry was performed with \texttt{hst1pass} \citep{anderson2022} on the flat-fielded calibrated images (FLT) downloaded from the MAST archive, i.e. processed through \texttt{calwf3} (v3.6.2).
The images were aligned to Gaia DR3 (\citealt{2023AA...674A...1G}) using the methodology described in \cite{2017wfc..rept...19B}. 
A second refining pass, aligning images to each other was also performed, to minimize the relative astrometric error. 
Both alignments were completed using the \texttt{DrizzlePac} function \texttt{TweakReg} \citep{2010bdrz.conf..382F}, with the catalogs from \texttt{hst1pass} as inputs.
The astrometrically aligned catalogs were then matched into a master catalog, containing records of stars that were measured in multiple exposures for each filter via functionality from the \texttt{astropy} package \citep{2018AJ....156..123A}. 

\begin{figure}[!b]
\centering
\includegraphics[width=0.9\linewidth]{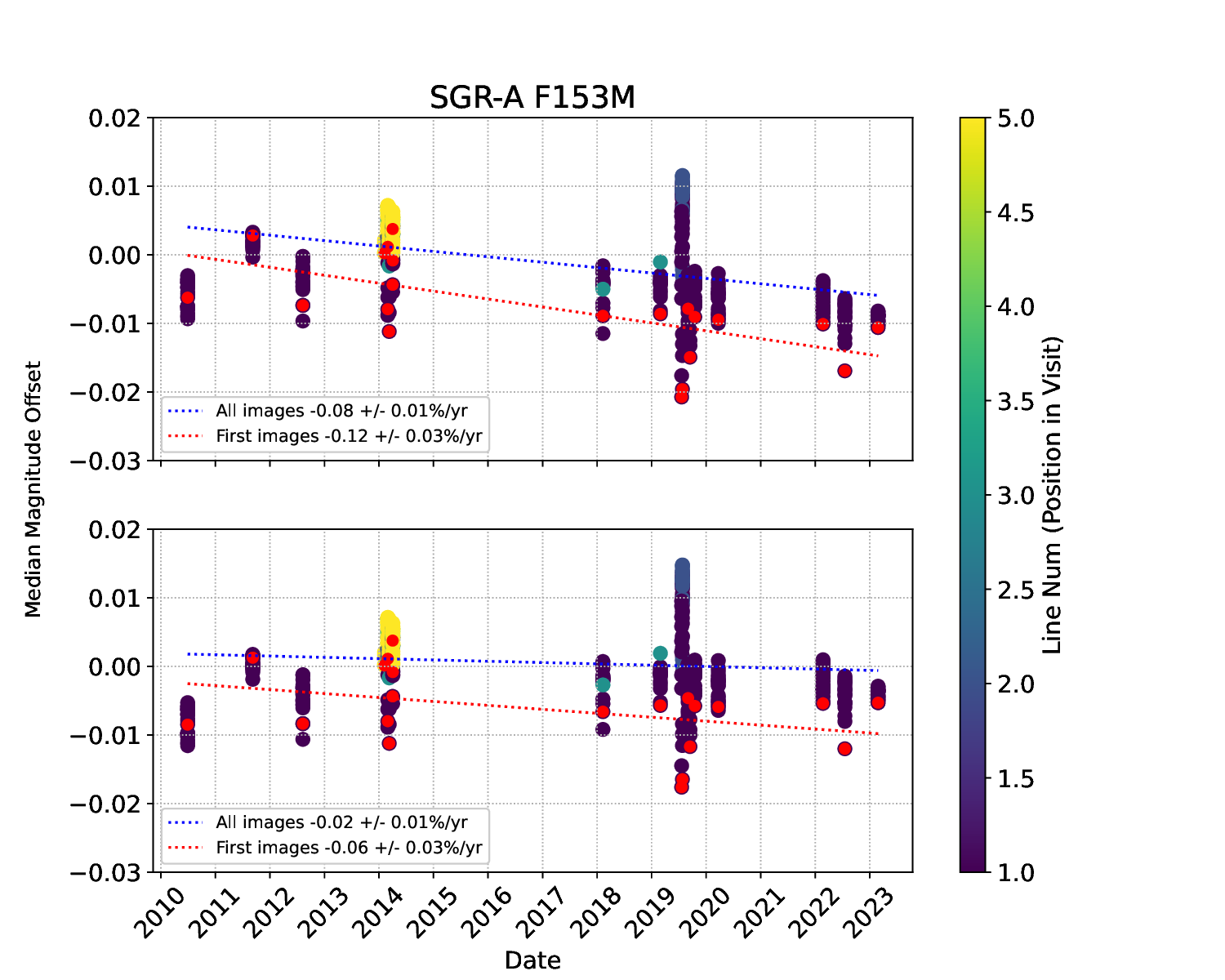}
\captionsetup{type=table, width=.9\linewidth}
\captionof{figure}{Median magnitude offset for $Sgr A^{*}$ stars measured in F153M images before (top) and after (bottom) applying the new time-dependent inverse sensitivities (\texttt{PHOTFLAM} values). The color of the points corresponds to the approximate position (\texttt{LINENUM}) in each visit when an exposure was taken.  Note that points colored in yellow correspond to exposures taken after many others in the same visit.  Due to the stellar crowding of this field, it is very likely these exposures are affected by self-persistence and thus show elevated count rates. Additionally, some visits observed very closely together in time (within hours) are also affected by persistence from preceding visits.
\label{fig:sgra}} 
\end{figure} 

Using only positions from the \texttt{hst1pass} catalogs, aperture photometry was performed on the stars using \texttt{photutils} version 1.8.0 \citep{larry_bradley_2021_5796924} and an implementation of the background measurements and error formulae described in \citep{davis1987}. 
While aperture photometry does make the absolute measurement of the star’s flux more vulnerable to crowding or blending, it removes systematic factors that can occur when fitting PSFs via \texttt{hst1pass}, and still allows the relative comparison of flux over time. 

We selected stars based on their brightness, photometric error, PSF fit quality (q parameter from \texttt{hst1pass}), and number of detections. Note that a q parameter value of 0 indicates that the profile of the star matches the PSF model exactly, while a q fit value of 1 indicates the model differs from the profile of the data by 100\% in each pixel.
Stars with a calibrated magnitude in the ST photometric system between 15.5 and 19 mag, photometric errors less than 0.065 mag, PSF fit quality $q <$ 0.65, and detected in 75\% of the total number of images were selected.  
These cuts ensure that only well-measured stars over a longer time-baseline are selected.   

\begin{figure}[!b]
\centering
\includegraphics[width=0.65\linewidth, angle=270]{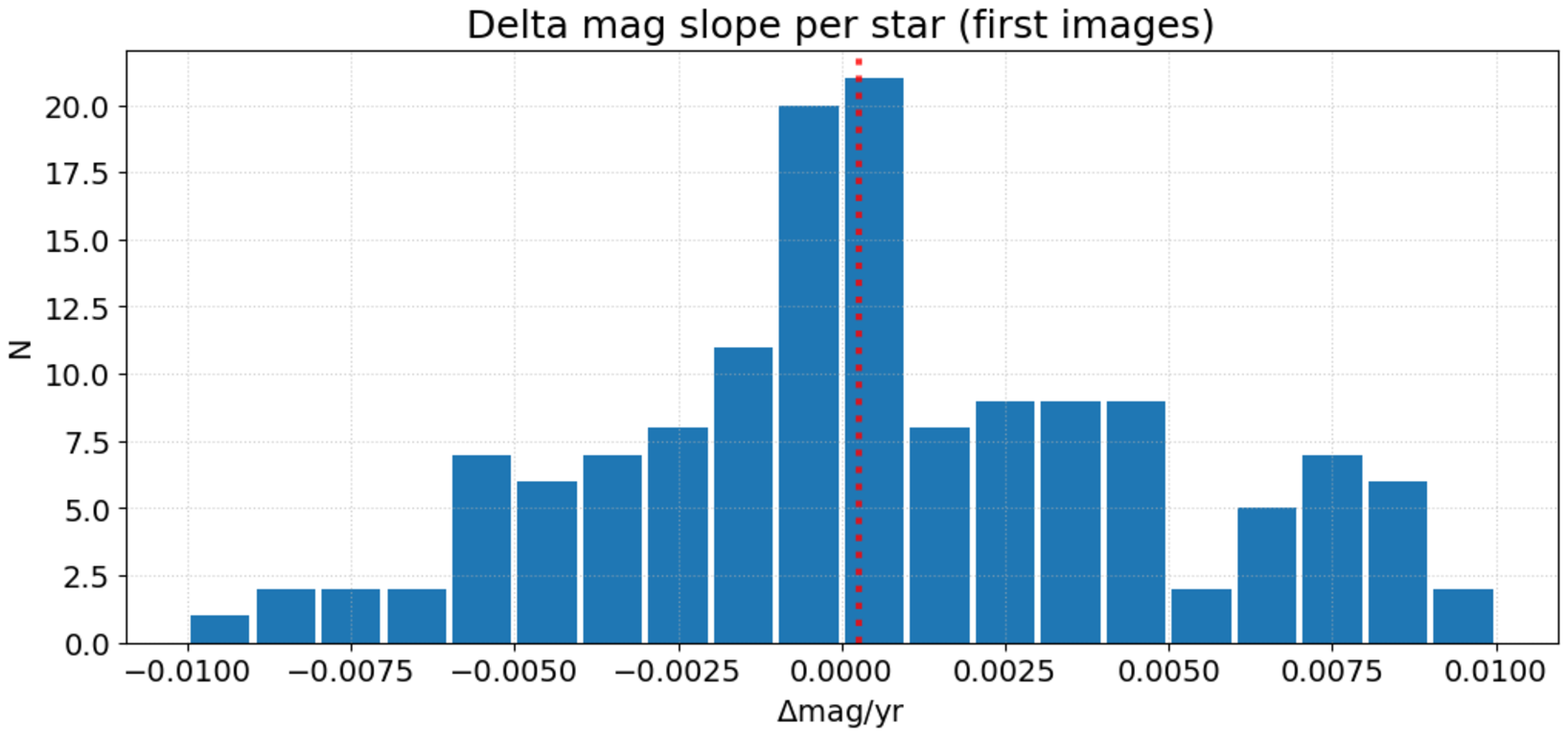}
\vspace{-2cm}
\captionsetup{type=table, width=.9\linewidth}
\captionof{figure}{Fitted slopes of selected star magnitudes in $Sgr A^{*}$ F153M images, after calibrating the observations with the new time-dependent inverse sensitivities (\texttt{PHOTFLAM} values). The median position is denoted by the vertical red line and is close to zero. Note that a positive value of $\Delta$mag$/$yr reflects increasing magnitude or loss of measured flux. \label{fig:histo_f153m}}
\end{figure} 

The F153M magnitudes calibrated with the \texttt{PHOTFLAM} value from the previous calibration (2020) compared to a reference magnitude (``Median Magnitude Offset'') are plotted as a function of time in the top panel of Fig.~\ref{fig:sgra}. 
The bottom panel shows the same plot for photometry performed on the FLT images created with the new \texttt{calwf3} pipeline (v3.7.2) and calibrated with the new time-dependent \texttt{PHOTFLAM} values, and so corrected for the time sensitivity changes of the detector. 
When fitting $Sgr A^{*}$ photometry calibrated with the constant \texttt{PHOTFLAM} value over the entire time range, a slope of -0.08$\pm$0.01 \%/yr is obtained, considering all measurements, or -0.11$\pm$0.03 \%/yr including only the first images observed in each visit. 
This distinction is useful since subsequent exposures in a orbit/visit might be increasingly affected by self-persistence \citep{2019wfc..rept....7B}, i.e. the observed region has a high-level of stellar crowding, resulting in increased dispersion of the measurements \citep{bajaj2022}. 
These slopes agree very well with the slope derived for the IR detector sensitivity's change in this wavelength regime by using multiple techniques, i.e. -0.060$\pm$0.005 (see Table~\ref{table:1}).
Note that the slopes calculated from the $Sgr A^{*}$ photometry are slightly biased towards steeper values as a result of the self- and external persistence present in this dataset.

\begin{figure}[!b]
\centering
\includegraphics[width=1.05\linewidth, angle=0]{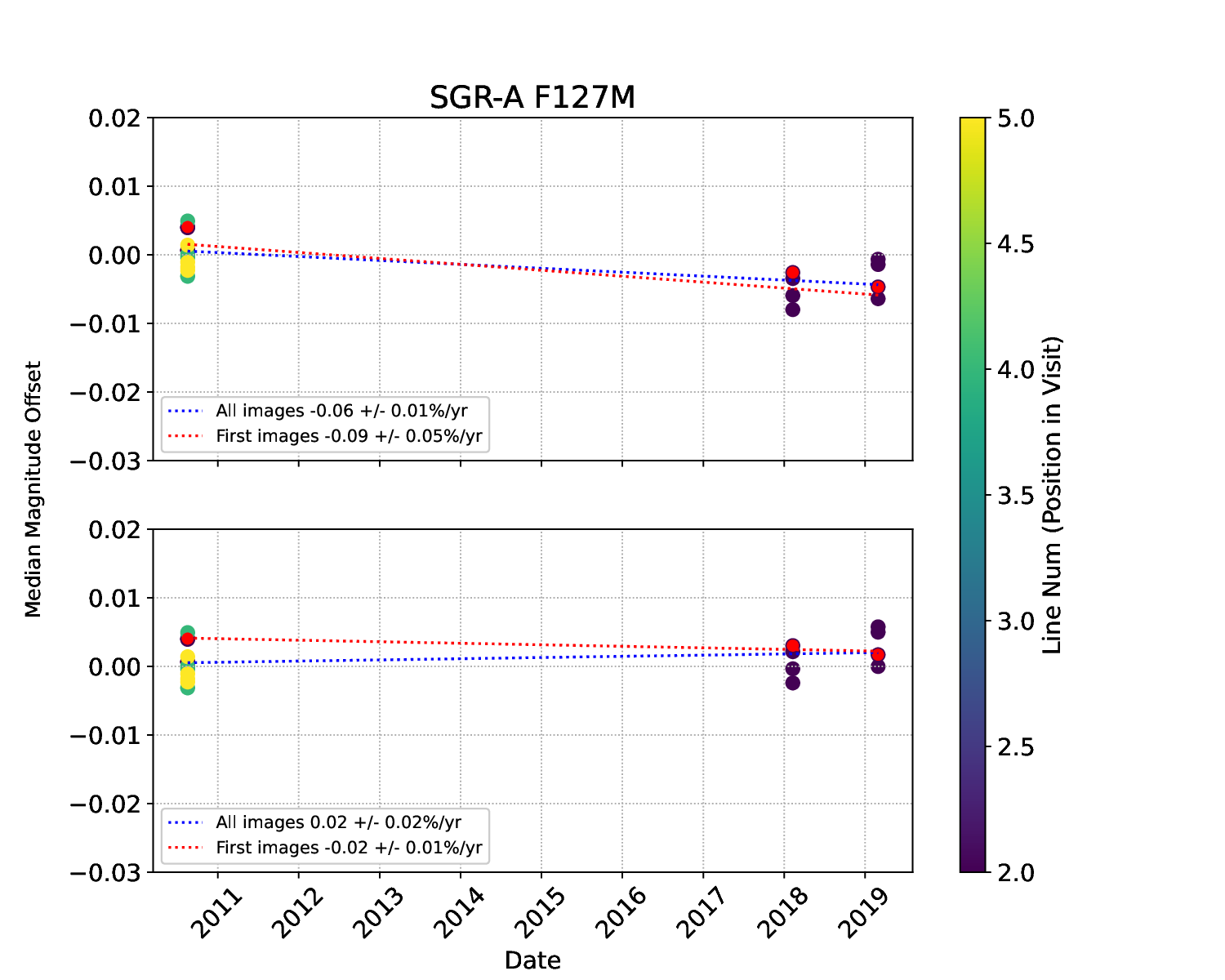}
\vspace{-0.8cm}
\captionsetup{type=table, width=.9\linewidth}
\captionof{figure}{Median magnitude offset for $Sgr A^{*}$ F127M images before (top) and after (bottom) applying the new time-dependent inverse sensitivities (\texttt{PHOTFLAM} values).
The color of the points corresponds to the approximate position (\texttt{LINENUM}) in each visit when an exposure was taken. There is only one early visit in this data, and the number of images differs between visits, so the blue slope (all images) is less well-constrained.
\label{fig:sgra_f127m}} 
\end{figure}

When processing the images with the new pipeline and calibrating the photometry with the time-dependent \texttt{PHOTFLAM} values, the measurements present a quite flat trend over the entire time range, with shallower slopes of -0.02$\pm$0.01\%/yr (or -0.05$\pm$0.03\%/yr, first images). 

In order to better quantify this effect, we also calculated the sensitivity change by measuring the change in count rate for individually selected stars.
For each star, a linear fit of its instrumental magnitude versus observation date (in years) was performed, resulting in a slope measuring its change in magnitude per year.
The median of these slopes was then computed to estimate the total change in sensitivity and is shown in the histogram of Fig.~\ref{fig:histo_f153m}, where a positive value of $\Delta$mag/yr reflects an increase in magnitude, i.e. a loss in flux over time. This plot shows a slight positive residual $\Delta$mag/yr, after calibrating the observations by using the new time-dependent \texttt{PHOTFLAM} values, supporting their validity. 

Fig.~\ref{fig:sgra_f127m} shows the same comparison as in Fig.~\ref{fig:sgra} but for photometry performed on F127M images of the same target.  
Stars with ST magnitudes between 16 and 20 mag, and cut with the same criterion used for the F153M observations, were selected for this analysis. 
Even with fewer observing epochs for this filter, the photometry corrected with the constant \texttt{PHOTFLAM} value from the previous calibraiton gives a slope of -0.06$\pm$0.01 \%/yr for all images, and -0.09$\pm$0.05 \%/yr for the first images, in very good agreement with the slopes calculated for this wavelength region (see Table~\ref{table:1}). 
The bottom panel of Fig.~\ref{fig:sgra_f127m} shows the same photometry calibrated with the new time-dependent \texttt{PHOTFLAM} values; the fitted slopes are now 0.02$\pm$0.02\%/yr (or -0.02$\pm$0.01\%/yr, first images), very nearly consistent with a flat trend.

\begin{figure}[!t]
\centering
\includegraphics[width=0.65\linewidth, angle=270]{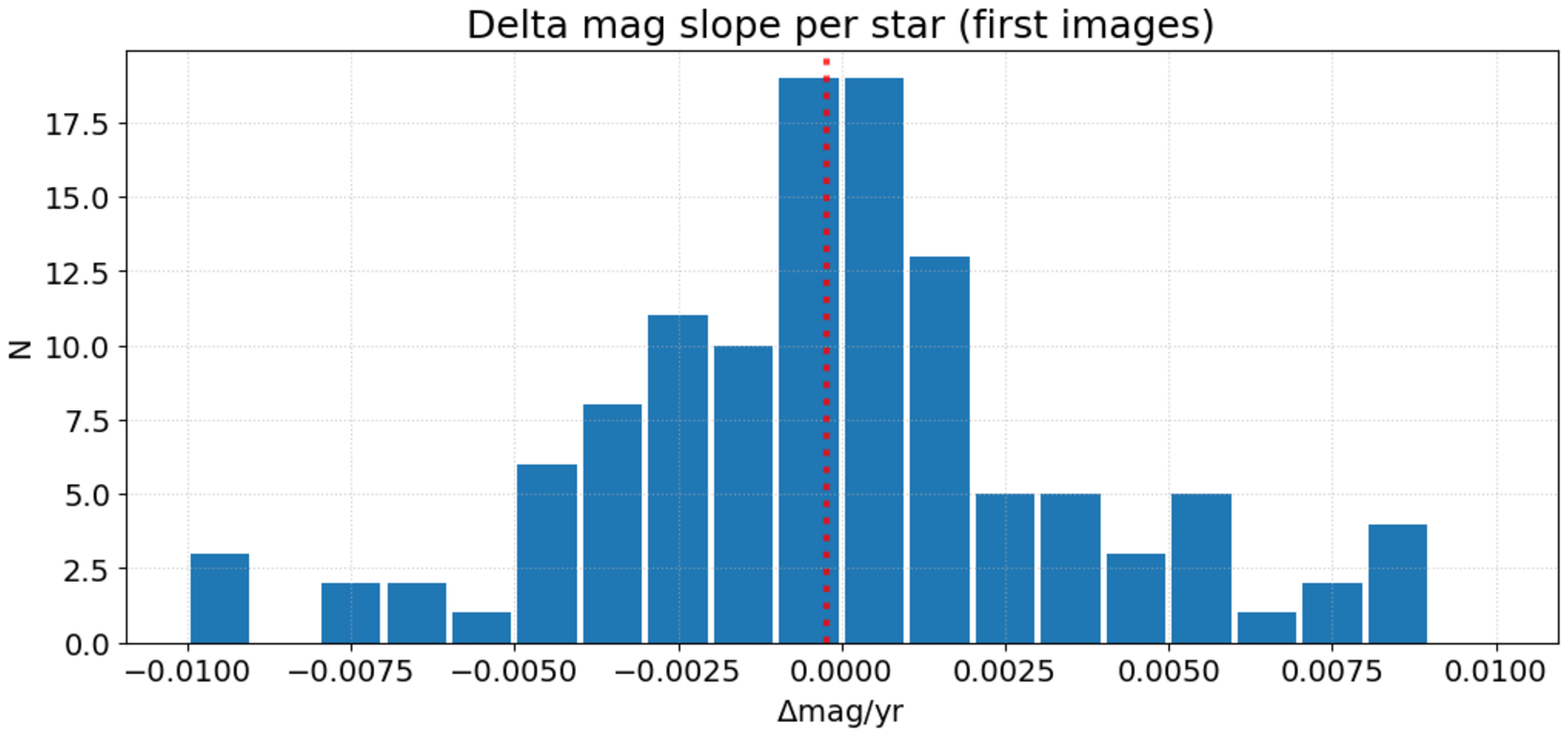}
\vspace{-1.5cm}
\caption{Same as Fig.~\ref{fig:histo_f153m} but for $Sgr A^{*}$ F127M images.}
\label{fig:histo_f127m}
\end{figure}

Fig.~\ref{fig:histo_f127m} shows the histogram of the median sensitivity change slopes measured for the F127M filter from selected stars in $Sgr A^{*}$. 
After correcting the photometry with the time-dependent \texttt{PHOTFLAM} values, the $\Delta$mag$/$yr has a slightly negative median, reflecting a slight decrease in magnitude with time, or a slight increase in flux, further supporting the previous result.

\vspace{-2cm}


\clearpage
\section{Recommendations for Users} \label{sec:recommendations}

It is important to note that updates to the inverse sensitivities do not impact the data (pixel values) of WFC3/IR images. 
The only change appears in the FLT image header, wherein the \texttt{PHOTFLAM} (erg$/$cm$^2/$\AA$/$e$^{-}$) and \texttt{PHOTFNU} (Jy$\times$s$/$e$^{-}$) keywords now reflect the value of those inverse sensitivities in the epoch recorded in the \texttt{EXPSTART} keyword. 
Therefore, in order to calibrate prior observations, it is only necessary to apply the new time-dependent inverse sensitivities to the flux measurements.

With the release of \texttt{calwf3} v3.7.2 in December 2024, the new values of \texttt{PHOTFLAM} and \texttt{PHOTFNU} will be populated in the FLT image headers for all data downloaded from MAST. 
Previously-retrieved data should either be re-downloaded to obtain the time-dependent inverse sensitivity header keywords or re-processed with \texttt{calwf3} v3.7.2 or higher.


If calculating inverse sensitivities using \texttt{stsynphot}, the new time-dependent filter throughput tables (see Table \ref{table:2}) need to be used. 
In order to derive the inverse sensitivity value at the desired epoch of observation, when creating the WFC3/IR bandpass object in \texttt{stsynphot}, the \verb+mjd#+ keyword must be used in the observation mode string (\texttt{obsmode}).\footnote{For \texttt{stsynphot} documentation regarding observation modes, see \href{https://stsynphot.readthedocs.io/en/latest/stsynphot/obsmode.html}{https://stsynphot.readthedocs.io/en/latest/} \href{https://stsynphot.readthedocs.io/en/latest/stsynphot/obsmode.html}{stsynphot/obsmode.html}. For explanation of the MJD keyword, see \href{https://stsynphot.readthedocs.io/en/latest/stsynphot/appendixb_specialkey.html}{https://stsynphot.readthedocs.io/en/latest/stsynphot/} \href{https://stsynphot.readthedocs.io/en/latest/stsynphot/appendixb_specialkey.html}{appendixb\_specialkey.html}.}

For example, a bandpass object used for calculating the F110W inverse sensitivity would previously be constructed as follows.

\begin{minipage}[!h]{0.9\linewidth} 

\vspace{2ex}
\begin{lstlisting}[language=Python]
import stsynphot as stsyn

bandpass = stsyn.band(obsmode='wfc3,ir,f110w')
\end{lstlisting}
\vspace{2ex}

\end{minipage}

In comparison, a F110W bandpass that is corrected for time-dependent sensitivity loss at MJD $= 57429$ would be constructed with the inclusion of a time-dependent value in the \texttt{obsmode} string. 


\begin{minipage}[!h]{0.9\linewidth} 

\vspace{2ex}

\begin{lstlisting}[language=Python]
bandpass = stsyn.band(obsmode='wfc3,ir,f110w,mjd#57429')
\end{lstlisting}
\vspace{2ex}

\end{minipage}

Additional instructions for calculating inverse sensitivities by using \texttt{stsynphot} and observation mode strings can be found in the \href{https://spacetelescope.github.io/hst_notebooks/notebooks/WFC3/zeropoints/zeropoints.html}{Calculating WFC3 Zeropoints with stsynphot} tutorial.\footnote{Zeropoints notebook: \href{https://spacetelescope.github.io/hst_notebooks/notebooks/WFC3/zeropoints/zeropoints.html}{https://spacetelescope.github.io/hst\_notebooks/notebooks/WFC3/zeropoints/zeropoints.html}}

\clearpage
\section{Conclusions \label{sec:conclusions}}

We derived new time-dependent inverse sensitivities for the WFC3/IR detector in all 15 filters, based on previously reported rates of sensitivity loss since launch \citepalias{2024wfc..rept....6M}.
The updated calibration provides internal photometric precision of $\lesssim$ 0.5\% for wide--, medium--, and narrow-band filters, an improvement compared to the previous calibration for which the precision across all filters was $\lesssim$ 0.6\%.

\medskip 

Changes in the calculation of these new inverse sensitivities, compared to the previous calibration values reported in 2020 \citep{bajaj2020}, can be summarized as follows:

\begin{itemize}
\item Time-dependent correction factors were calculated after establishing wavelength-dependent sensitivity loss rates using multiple methods and targets (open cluster spatial scans, staring mode observations of globular clusters, grism observations of standard stars);
\item Staring mode standard star photometry was corrected for sensitivity changes by applying the aforementioned correction factors before calculating the inverse sensitivities;
\item Four additional years (2019 -- 2023) of standard star photometry were included when calculating the inverse sensitivities;
\item The most updated version of the CALSPEC SEDs for the five standard stars were used.
\end{itemize}

A new IMPHTTAB reference file was created and then tested on observations of the CALSPEC standard star GD153 and for a region towards the Galactic center ($Sgr A^{*}$) collected over a time interval $\ge$ 10 years. 
In both cases, aperture photometry corrected by using the time-dependent \texttt{PHOTFLAM} values, populated in the header of the images processed with the new version of \texttt{calwf3} (v3.7.2), shows a decreased dispersion of the measurements and a flattening trend over time. 

The IMPHTTAB \footnote{Updated image photometry lookup table reference file (IMPHTTAB): \texttt{8bq13281i\_imp.fits}} was delivered to the \href{https://hst-crds.stsci.edu/}{HST Calibration Reference Data System} (CRDS),\footnote{HST Calibration Reference Data System: \href{https://hst-crds.stsci.edu/}{https://hst-crds.stsci.edu/}} and all WFC3/IR data are expected to be reprocessed in the MAST archive with the updated \texttt{calwf3} pipeline in late 2024.
Time-dependent photometry keyword values (\texttt{PHOTFLAM} and \texttt{PHOTNU}) will populate the headers in FLT images following MAST reprocessing, or FLT images previously downloaded \textbf{and} then re-processed with \texttt{calwf3} v3.7.2. 
In the latter case, we highly recommend the user confirm that the \texttt{CAL\_VER} keyword in the header of the previously-downloaded, reprocessed images is set to a version of \texttt{calwf3} that is compatible with the new time-dependent IMPHTTAB reference file ($\geq$ 3.7.2). 

New filter throughput curves for all filters have also been delivered to CRDS so that they can be used in \texttt{stsynphot} simulations; see Table~\ref{table:2} for a list of file names. 


The \href{https://www.stsci.edu/hst/instrumentation/wfc3/data-analysis/photometric-calibration/ir-photometric-calibration}{WFC3/IR photometric calibration web page} \footnote{IR Photometric Calibration: \href{https://www.stsci.edu/hst/instrumentation/wfc3/data-analysis/photometric-calibration/ir-photometric-calibration}{https://www.stsci.edu/hst/instrumentation/wfc3/data-analysis/photometric-calibration/} \href{https://www.stsci.edu/hst/instrumentation/wfc3/data-analysis/photometric-calibration/ir-photometric-calibration}{ir-photometric-calibration}} provides a table with the new inverse sensitivity values (calculated at the reference epoch, MJD = 55008), and the corresponding ZPs in different photometric systems, namely ST, AB, and Vega.

\section*{Acknowledgements}
This work was accomplished with the contribution of all members of the WFC3 Photometry Group, including Ky Huynh, Sachindev Shenoy, and Debopam Som.
We would also like to thank Ralph Bohlin for his insightful comments on the absolute flux calibration, and Isabel Rivera for delivering the new IMPHTTAB reference file and filter throughputs.
We also thank the reviewer of this document, Mitchell Revalski, and Joel Green for a through editorial review.

\clearpage
\addcontentsline{toc}{section}{References}

\raggedright
\bibliography{ref.bib}

\begin{thebibliography}{}
\expandafter\ifx\csname natexlab\endcsname\relax\def\natexlab#1{#1}\fi
\providecommand{\url}[1]{\href{#1}{#1}}
\providecommand{\dodoi}[1]{doi:~\href{http://doi.org/#1}{\nolinkurl{#1}}}
\providecommand{\doeprint}[1]{\href{http://ascl.net/#1}{\nolinkurl{http://ascl.net/#1}}}
\providecommand{\doarXiv}[1]{\href{https://arxiv.org/abs/#1}{\nolinkurl{https://arxiv.org/abs/#1}}}

\bibitem[{{Anderson}(2022)}]{anderson2022}
{Anderson}, J. 2022, {One-Pass HST Photometry with \texttt{hst1pass}}, WFC3 Instrument Science Report 2022-05, 55 pages

\bibitem[{{Astropy Collaboration} {et~al.}(2013{\natexlab{a}}){Astropy Collaboration}, {Robitaille}, {Tollerud}, {Greenfield}, {Droettboom}, {Bray}, {Aldcroft}, {Davis}, {Ginsburg}, {Price-Whelan}, {Kerzendorf}, {Conley}, {Crighton}, {Barbary}, {Muna}, {Ferguson}, {Grollier}, {Parikh}, {Nair}, {Unther}, {Deil}, {Woillez}, {Conseil}, {Kramer}, {Turner}, {Singer}, {Fox}, {Weaver}, {Zabalza}, {Edwards}, {Azalee Bostroem}, {Burke}, {Casey}, {Crawford}, {Dencheva}, {Ely}, {Jenness}, {Labrie}, {Lim}, {Pierfederici}, {Pontzen}, {Ptak}, {Refsdal}, {Servillat}, \& {Streicher}}]{2013A&A...558A..33A}
{Astropy Collaboration}, {Robitaille}, T.~P., {Tollerud}, E.~J., {et~al.} 2013{\natexlab{a}}, {Astropy: A community Python package for astronomy}, \aap, 558, A33, \dodoi{10.1051/0004-6361/201322068}

\bibitem[{{Astropy Collaboration} {et~al.}(2013{\natexlab{b}}){Astropy Collaboration}, {Robitaille}, {Tollerud}, {Greenfield}, {Droettboom}, {Bray}, {Aldcroft}, {Davis}, {Ginsburg}, {Price-Whelan}, {Kerzendorf}, {Conley}, {Crighton}, {Barbary}, {Muna}, {Ferguson}, {Grollier}, {Parikh}, {Nair}, {Unther}, {Deil}, {Woillez}, {Conseil}, {Kramer}, {Turner}, {Singer}, {Fox}, {Weaver}, {Zabalza}, {Edwards}, {Azalee Bostroem}, {Burke}, {Casey}, {Crawford}, {Dencheva}, {Ely}, {Jenness}, {Labrie}, {Lim}, {Pierfederici}, {Pontzen}, {Ptak}, {Refsdal}, {Servillat}, \& {Streicher}}]{2013AA...558A..33A}
---. 2013{\natexlab{b}}, {Astropy: A community Python package for astronomy}, \aap, 558, A33, \dodoi{10.1051/0004-6361/201322068}

\bibitem[{{Astropy Collaboration} {et~al.}(2018){Astropy Collaboration}, {Price-Whelan}, {Sip{\H{o}}cz}, {G{\"u}nther}, {Lim}, {Crawford}, {Conseil}, {Shupe}, {Craig}, {Dencheva}, {Ginsburg}, {VanderPlas}, {Bradley}, {P{\'e}rez-Su{\'a}rez}, {de Val-Borro}, {Aldcroft}, {Cruz}, {Robitaille}, {Tollerud}, {Ardelean}, {Babej}, {Bach}, {Bachetti}, {Bakanov}, {Bamford}, {Barentsen}, {Barmby}, {Baumbach}, {Berry}, {Biscani}, {Boquien}, {Bostroem}, {Bouma}, {Brammer}, {Bray}, {Breytenbach}, {Buddelmeijer}, {Burke}, {Calderone}, {Cano Rodr{\'\i}guez}, {Cara}, {Cardoso}, {Cheedella}, {Copin}, {Corrales}, {Crichton}, {D'Avella}, {Deil}, {Depagne}, {Dietrich}, {Donath}, {Droettboom}, {Earl}, {Erben}, {Fabbro}, {Ferreira}, {Finethy}, {Fox}, {Garrison}, {Gibbons}, {Goldstein}, {Gommers}, {Greco}, {Greenfield}, {Groener}, {Grollier}, {Hagen}, {Hirst}, {Homeier}, {Horton}, {Hosseinzadeh}, {Hu}, {Hunkeler}, {Ivezi{\'c}}, {Jain}, {Jenness}, {Kanarek}, {Kendrew}, {Kern}, {Kerzendorf}, {Khvalko}, {King}, {Kirkby}, {Kulkarni},
  {Kumar}, {Lee}, {Lenz}, {Littlefair}, {Ma}, {Macleod}, {Mastropietro}, {McCully}, {Montagnac}, {Morris}, {Mueller}, {Mumford}, {Muna}, {Murphy}, {Nelson}, {Nguyen}, {Ninan}, {N{\"o}the}, {Ogaz}, {Oh}, {Parejko}, {Parley}, {Pascual}, {Patil}, {Patil}, {Plunkett}, {Prochaska}, {Rastogi}, {Reddy Janga}, {Sabater}, {Sakurikar}, {Seifert}, {Sherbert}, {Sherwood-Taylor}, {Shih}, {Sick}, {Silbiger}, {Singanamalla}, {Singer}, {Sladen}, {Sooley}, {Sornarajah}, {Streicher}, {Teuben}, {Thomas}, {Tremblay}, {Turner}, {Terr{\'o}n}, {van Kerkwijk}, {de la Vega}, {Watkins}, {Weaver}, {Whitmore}, {Woillez}, {Zabalza}, \& {Astropy Contributors}}]{2018AJ....156..123A}
{Astropy Collaboration}, {Price-Whelan}, A.~M., {Sip{\H{o}}cz}, B.~M., {et~al.} 2018, {The Astropy Project: Building an Open-science Project and Status of the v2.0 Core Package}, \aj, 156, 123, \dodoi{10.3847/1538-3881/aabc4f}

\bibitem[{{Bajaj}(2017)}]{2017wfc..rept...19B}
{Bajaj}, V. 2017, {Aligning HST Images to Gaia: A Faster Mosaicking Workflow}, WFC3 Instrument Science Report 2017-19, 4 pages

\bibitem[{{Bajaj}(2019)}]{2019wfc..rept....7B}
---. 2019, {WFC3/IR Photometric Repeatability}, WFC3 Instrument Science Report 2019-07

\bibitem[{{Bajaj} {et~al.}(2020){Bajaj}, {Calamida}, \& {Mack}}]{bajaj2020}
{Bajaj}, V., {Calamida}, A., \& {Mack}, J. 2020, {Updated WFC3/IR Photometric Calibration}, WFC3 Instrument Science Report 2020-10

\bibitem[{{Bajaj} {et~al.}(2022){Bajaj}, {Calamida}, {Mack}, \& {Som}}]{bajaj2022}
{Bajaj}, V., {Calamida}, A., {Mack}, J., \& {Som}, D. 2022, {WFC3/IR Photometric Stability Stellar Cluster Study}, WFC3 Instrument Science Report 2022-07, 28 pages

\bibitem[{{Bohlin} {et~al.}(2020){Bohlin}, {Hubeny}, \& {Rauch}}]{bohlin2020}
{Bohlin}, R.~C., {Hubeny}, I., \& {Rauch}, T. 2020, {New Grids of Pure-hydrogen White Dwarf NLTE Model Atmospheres and the HST/STIS Flux Calibration}, \aj, 160, 21, \dodoi{10.3847/1538-3881/ab94b4}

\bibitem[{{Bohlin} \& {Lockwood}(2022)}]{bohlin2022}
{Bohlin}, R.~C., \& {Lockwood}, S. 2022, {Update of the STIS CTE Correction Formula for Stellar Spectra}, STIS Instrument Science Report 2022-07, 11 pages

\bibitem[{Bradley {et~al.}(2021)Bradley, Sipőcz, Robitaille, Tollerud, Vinícius, Deil, Barbary, Wilson, Busko, Donath, Günther, Cara, krachyon, Conseil, Bostroem, Droettboom, Bray, Lim, Bratholm, Barentsen, Craig, Rathi, Pascual, Perren, Georgiev, de~Val-Borro, Kerzendorf, Bach, Quint, \& Souchereau}]{larry_bradley_2021_5796924}
Bradley, L., Sipőcz, B., Robitaille, T., {et~al.} 2021, astropy/photutils: 1.3.0, 1.3.0,  Zenodo, \dodoi{10.5281/zenodo.5796924}

\bibitem[{{Calamida} {et~al.}(2021){Calamida}, {Mack}, {Medina}, {Shanahan}, {Bajaj}, \& {Deustua}}]{calamida2021}
{Calamida}, A., {Mack}, J., {Medina}, J., {et~al.} 2021, {New time-dependent WFC3 UVIS inverse sensitivities}, WFC3 Instrument Science Report 2021-04

\bibitem[{{Calamida} {et~al.}(2022){Calamida}, {Bajaj}, {Mack}, {Marinelli}, {Medina}, {Pidgeon}, {Kozhurina-Platais}, {Shanahan}, \& {Som}}]{calamida2022}
{Calamida}, A., {Bajaj}, V., {Mack}, J., {et~al.} 2022, {New Photometric Calibration of the Wide Field Camera 3 Detectors}, \aj, 164, 32, \dodoi{10.3847/1538-3881/ac73f0}

\bibitem[{{Davis}(1987)}]{davis1987}
{Davis}, L. 1987, {Specifications for the Aperture Photometry Package}, Tech. rep., National Optical Astronomy Observatories.
\newblock \url{https://iraf.net/irafdocs/apspec.pdf}

\bibitem[{{Fruchter} \& {et al.}(2010)}]{2010bdrz.conf..382F}
{Fruchter}, A.~S., \& {et al.} 2010, in 2010 Space Telescope Science Institute Calibration Workshop, 382--387

\bibitem[{{Gaia Collaboration} {et~al.}(2016){Gaia Collaboration}, {Prusti}, {de Bruijne}, {Brown}, {Vallenari}, {Babusiaux}, {Bailer-Jones}, {Bastian}, {Biermann}, {Evans}, {Eyer}, {Jansen}, {Jordi}, {Klioner}, {Lammers}, {Lindegren}, {Luri}, {Mignard}, {Milligan}, {Panem}, {Poinsignon}, {Pourbaix}, {Randich}, {Sarri}, {Sartoretti}, {Siddiqui}, {Soubiran}, {Valette}, {van Leeuwen}, {Walton}, {Aerts}, {Arenou}, {Cropper}, {Drimmel}, {H{\o}g}, {Katz}, {Lattanzi}, {O'Mullane}, {Grebel}, {Holland}, {Huc}, {Passot}, {Bramante}, {Cacciari}, {Casta{\~n}eda}, {Chaoul}, {Cheek}, {De Angeli}, {Fabricius}, {Guerra}, {Hern{\'a}ndez}, {Jean-Antoine-Piccolo}, {Masana}, {Messineo}, {Mowlavi}, {Nienartowicz}, {Ord{\'o}{\~n}ez-Blanco}, {Panuzzo}, {Portell}, {Richards}, {Riello}, {Seabroke}, {Tanga}, {Th{\'e}venin}, {Torra}, {Els}, {Gracia-Abril}, {Comoretto}, {Garcia-Reinaldos}, {Lock}, {Mercier}, {Altmann}, {Andrae}, {Astraatmadja}, {Bellas-Velidis}, {Benson}, {Berthier}, {Blomme}, {Busso}, {Carry}, {Cellino}, {Clementini},
  {Cowell}, {Creevey}, {Cuypers}, {Davidson}, {De Ridder}, {de Torres}, {Delchambre}, {Dell'Oro}, {Ducourant}, {Fr{\'e}mat}, {Garc{\'\i}a-Torres}, {Gosset}, {Halbwachs}, {Hambly}, {Harrison}, {Hauser}, {Hestroffer}, {Hodgkin}, {Huckle}, {Hutton}, {Jasniewicz}, {Jordan}, {Kontizas}, {Korn}, {Lanzafame}, {Manteiga}, {Moitinho}, {Muinonen}, {Osinde}, {Pancino}, {Pauwels}, {Petit}, {Recio-Blanco}, {Robin}, {Sarro}, {Siopis}, {Smith}, {Smith}, {Sozzetti}, {Thuillot}, {van Reeven}, {Viala}, {Abbas}, {Abreu Aramburu}, {Accart}, {Aguado}, {Allan}, {Allasia}, {Altavilla}, {{\'A}lvarez}, {Alves}, {Anderson}, {Andrei}, {Anglada Varela}, {Antiche}, {Antoja}, {Ant{\'o}n}, {Arcay}, {Atzei}, {Ayache}, {Bach}, {Baker}, {Balaguer-N{\'u}{\~n}ez}, {Barache}, {Barata}, {Barbier}, {Barblan}, {Baroni}, {Barrado y Navascu{\'e}s}, {Barros}, {Barstow}, {Becciani}, {Bellazzini}, {Bellei}, {Bello Garc{\'\i}a}, {Belokurov}, {Bendjoya}, {Berihuete}, {Bianchi}, {Bienaym{\'e}}, {Billebaud}, {Blagorodnova}, {Blanco-Cuaresma}, {Boch},
  {Bombrun}, {Borrachero}, {Bouquillon}, {Bourda}, {Bouy}, {Bragaglia}, {Breddels}, {Brouillet}, {Br{\"u}semeister}, {Bucciarelli}, {Budnik}, {Burgess}, {Burgon}, {Burlacu}, {Busonero}, {Buzzi}, {Caffau}, {Cambras}, {Campbell}, {Cancelliere}, {Cantat-Gaudin}, {Carlucci}, {Carrasco}, {Castellani}, {Charlot}, {Charnas}, {Charvet}, {Chassat}, {Chiavassa}, {Clotet}, {Cocozza}, {Collins}, {Collins}, {Costigan}, {Crifo}, {Cross}, {Crosta}, {Crowley}, {Dafonte}, {Damerdji}, {Dapergolas}, {David}, {David}, {De Cat}, {de Felice}, {de Laverny}, {De Luise}, {De March}, {de Martino}, {de Souza}, {Debosscher}, {del Pozo}, {Delbo}, {Delgado}, {Delgado}, {di Marco}, {Di Matteo}, {Diakite}, {Distefano}, {Dolding}, {Dos Anjos}, {Drazinos}, {Dur{\'a}n}, {Dzigan}, {Ecale}, {Edvardsson}, {Enke}, {Erdmann}, {Escolar}, {Espina}, {Evans}, {Eynard Bontemps}, {Fabre}, {Fabrizio}, {Faigler}, {Falc{\~a}o}, {Farr{\`a}s Casas}, {Faye}, {Federici}, {Fedorets}, {Fern{\'a}ndez-Hern{\'a}ndez}, {Fernique}, {Fienga}, {Figueras}, {Filippi},
  {Findeisen}, {Fonti}, {Fouesneau}, {Fraile}, {Fraser}, {Fuchs}, {Furnell}, {Gai}, {Galleti}, {Galluccio}, {Garabato}, {Garc{\'\i}a-Sedano}, {Gar{\'e}}, {Garofalo}, {Garralda}, {Gavras}, {Gerssen}, {Geyer}, {Gilmore}, {Girona}, {Giuffrida}, {Gomes}, {Gonz{\'a}lez-Marcos}, {Gonz{\'a}lez-N{\'u}{\~n}ez}, {Gonz{\'a}lez-Vidal}, {Granvik}, {Guerrier}, {Guillout}, {Guiraud}, {G{\'u}rpide}, {Guti{\'e}rrez-S{\'a}nchez}, {Guy}, {Haigron}, {Hatzidimitriou}, {Haywood}, {Heiter}, {Helmi}, {Hobbs}, {Hofmann}, {Holl}, {Holland}, {Hunt}, {Hypki}, {Icardi}, {Irwin}, {Jevardat de Fombelle}, {Jofr{\'e}}, {Jonker}, {Jorissen}, {Julbe}, {Karampelas}, {Kochoska}, {Kohley}, {Kolenberg}, {Kontizas}, {Koposov}, {Kordopatis}, {Koubsky}, {Kowalczyk}, {Krone-Martins}, {Kudryashova}, {Kull}, {Bachchan}, {Lacoste-Seris}, {Lanza}, {Lavigne}, {Le Poncin-Lafitte}, {Lebreton}, {Lebzelter}, {Leccia}, {Leclerc}, {Lecoeur-Taibi}, {Lemaitre}, {Lenhardt}, {Leroux}, {Liao}, {Licata}, {Lindstr{\o}m}, {Lister}, {Livanou}, {Lobel}, {L{\"o}ffler},
  {L{\'o}pez}, {Lopez-Lozano}, {Lorenz}, {Loureiro}, {MacDonald}, {Magalh{\~a}es Fernandes}, {Managau}, {Mann}, {Mantelet}, {Marchal}, {Marchant}, {Marconi}, {Marie}, {Marinoni}, {Marrese}, {Marschalk{\'o}}, {Marshall}, {Mart{\'\i}n-Fleitas}, {Martino}, {Mary}, {Matijevi{\v{c}}}, {Mazeh}, {McMillan}, {Messina}, {Mestre}, {Michalik}, {Millar}, {Miranda}, {Molina}, {Molinaro}, {Molinaro}, {Moln{\'a}r}, {Moniez}, {Montegriffo}, {Monteiro}, {Mor}, {Mora}, {Morbidelli}, {Morel}, {Morgenthaler}, {Morley}, {Morris}, {Mulone}, {Muraveva}, {Musella}, {Narbonne}, {Nelemans}, {Nicastro}, {Noval}, {Ord{\'e}novic}, {Ordieres-Mer{\'e}}, {Osborne}, {Pagani}, {Pagano}, {Pailler}, {Palacin}, {Palaversa}, {Parsons}, {Paulsen}, {Pecoraro}, {Pedrosa}, {Pentik{\"a}inen}, {Pereira}, {Pichon}, {Piersimoni}, {Pineau}, {Plachy}, {Plum}, {Poujoulet}, {Pr{\v{s}}a}, {Pulone}, {Ragaini}, {Rago}, {Rambaux}, {Ramos-Lerate}, {Ranalli}, {Rauw}, {Read}, {Regibo}, {Renk}, {Reyl{\'e}}, {Ribeiro}, {Rimoldini}, {Ripepi}, {Riva}, {Rixon},
  {Roelens}, {Romero-G{\'o}mez}, {Rowell}, {Royer}, {Rudolph}, {Ruiz-Dern}, {Sadowski}, {Sagrist{\`a} Sell{\'e}s}, {Sahlmann}, {Salgado}, {Salguero}, {Sarasso}, {Savietto}, {Schnorhk}, {Schultheis}, {Sciacca}, {Segol}, {Segovia}, {Segransan}, {Serpell}, {Shih}, {Smareglia}, {Smart}, {Smith}, {Solano}, {Solitro}, {Sordo}, {Soria Nieto}, {Souchay}, {Spagna}, {Spoto}, {Stampa}, {Steele}, {Steidelm{\"u}ller}, {Stephenson}, {Stoev}, {Suess}, {S{\"u}veges}, {Surdej}, {Szabados}, {Szegedi-Elek}, {Tapiador}, {Taris}, {Tauran}, {Taylor}, {Teixeira}, {Terrett}, {Tingley}, {Trager}, {Turon}, {Ulla}, {Utrilla}, {Valentini}, {van Elteren}, {Van Hemelryck}, {van Leeuwen}, {Varadi}, {Vecchiato}, {Veljanoski}, {Via}, {Vicente}, {Vogt}, {Voss}, {Votruba}, {Voutsinas}, {Walmsley}, {Weiler}, {Weingrill}, {Werner}, {Wevers}, {Whitehead}, {Wyrzykowski}, {Yoldas}, {{\v{Z}}erjal}, {Zucker}, {Zurbach}, {Zwitter}, {Alecu}, {Allen}, {Allende Prieto}, {Amorim}, {Anglada-Escud{\'e}}, {Arsenijevic}, {Azaz}, {Balm}, {Beck}, {Bernstein},
  {Bigot}, {Bijaoui}, {Blasco}, {Bonfigli}, {Bono}, {Boudreault}, {Bressan}, {Brown}, {Brunet}, {Bunclark}, {Buonanno}, {Butkevich}, {Carret}, {Carrion}, {Chemin}, {Ch{\'e}reau}, {Corcione}, {Darmigny}, {de Boer}, {de Teodoro}, {de Zeeuw}, {Delle Luche}, {Domingues}, {Dubath}, {Fodor}, {Fr{\'e}zouls}, {Fries}, {Fustes}, {Fyfe}, {Gallardo}, {Gallegos}, {Gardiol}, {Gebran}, {Gomboc}, {G{\'o}mez}, {Grux}, {Gueguen}, {Heyrovsky}, {Hoar}, {Iannicola}, {Isasi Parache}, {Janotto}, {Joliet}, {Jonckheere}, {Keil}, {Kim}, {Klagyivik}, {Klar}, {Knude}, {Kochukhov}, {Kolka}, {Kos}, {Kutka}, {Lainey}, {LeBouquin}, {Liu}, {Loreggia}, {Makarov}, {Marseille}, {Martayan}, {Martinez-Rubi}, {Massart}, {Meynadier}, {Mignot}, {Munari}, {Nguyen}, {Nordlander}, {Ocvirk}, {O'Flaherty}, {Olias Sanz}, {Ortiz}, {Osorio}, {Oszkiewicz}, {Ouzounis}, {Palmer}, {Park}, {Pasquato}, {Peltzer}, {Peralta}, {P{\'e}turaud}, {Pieniluoma}, {Pigozzi}, {Poels}, {Prat}, {Prod'homme}, {Raison}, {Rebordao}, {Risquez}, {Rocca-Volmerange}, {Rosen},
  {Ruiz-Fuertes}, {Russo}, {Sembay}, {Serraller Vizcaino}, {Short}, {Siebert}, {Silva}, {Sinachopoulos}, {Slezak}, {Soffel}, {Sosnowska}, {Strai{\v{z}}ys}, {ter Linden}, {Terrell}, {Theil}, {Tiede}, {Troisi}, {Tsalmantza}, {Tur}, {Vaccari}, {Vachier}, {Valles}, {Van Hamme}, {Veltz}, {Virtanen}, {Wallut}, {Wichmann}, {Wilkinson}, {Ziaeepour}, \& {Zschocke}}]{2016AA...595A...1G}
{Gaia Collaboration}, {Prusti}, T., {de Bruijne}, J.~H.~J., {et~al.} 2016, {The Gaia mission}, \aap, 595, A1, \dodoi{10.1051/0004-6361/201629272}

\bibitem[{{Gaia Collaboration} {et~al.}(2023){Gaia Collaboration}, {Vallenari}, {Brown}, {Prusti}, {de Bruijne}, {Arenou}, {Babusiaux}, {Biermann}, {Creevey}, {Ducourant}, {Evans}, {Eyer}, {Guerra}, {Hutton}, {Jordi}, {Klioner}, {Lammers}, {Lindegren}, {Luri}, {Mignard}, {Panem}, {Pourbaix}, {Randich}, {Sartoretti}, {Soubiran}, {Tanga}, {Walton}, {Bailer-Jones}, {Bastian}, {Drimmel}, {Jansen}, {Katz}, {Lattanzi}, {van Leeuwen}, {Bakker}, {Cacciari}, {Casta{\~n}eda}, {De Angeli}, {Fabricius}, {Fouesneau}, {Fr{\'e}mat}, {Galluccio}, {Guerrier}, {Heiter}, {Masana}, {Messineo}, {Mowlavi}, {Nicolas}, {Nienartowicz}, {Pailler}, {Panuzzo}, {Riclet}, {Roux}, {Seabroke}, {Sordo}, {Th{\'e}venin}, {Gracia-Abril}, {Portell}, {Teyssier}, {Altmann}, {Andrae}, {Audard}, {Bellas-Velidis}, {Benson}, {Berthier}, {Blomme}, {Burgess}, {Busonero}, {Busso}, {C{\'a}novas}, {Carry}, {Cellino}, {Cheek}, {Clementini}, {Damerdji}, {Davidson}, {de Teodoro}, {Nu{\~n}ez Campos}, {Delchambre}, {Dell'Oro}, {Esquej},
  {Fern{\'a}ndez-Hern{\'a}ndez}, {Fraile}, {Garabato}, {Garc{\'\i}a-Lario}, {Gosset}, {Haigron}, {Halbwachs}, {Hambly}, {Harrison}, {Hern{\'a}ndez}, {Hestroffer}, {Hodgkin}, {Holl}, {Jan{\ss}en}, {Jevardat de Fombelle}, {Jordan}, {Krone-Martins}, {Lanzafame}, {L{\"o}ffler}, {Marchal}, {Marrese}, {Moitinho}, {Muinonen}, {Osborne}, {Pancino}, {Pauwels}, {Recio-Blanco}, {Reyl{\'e}}, {Riello}, {Rimoldini}, {Roegiers}, {Rybizki}, {Sarro}, {Siopis}, {Smith}, {Sozzetti}, {Utrilla}, {van Leeuwen}, {Abbas}, {{\'A}brah{\'a}m}, {Abreu Aramburu}, {Aerts}, {Aguado}, {Ajaj}, {Aldea-Montero}, {Altavilla}, {{\'A}lvarez}, {Alves}, {Anders}, {Anderson}, {Anglada Varela}, {Antoja}, {Baines}, {Baker}, {Balaguer-N{\'u}{\~n}ez}, {Balbinot}, {Balog}, {Barache}, {Barbato}, {Barros}, {Barstow}, {Bartolom{\'e}}, {Bassilana}, {Bauchet}, {Becciani}, {Bellazzini}, {Berihuete}, {Bernet}, {Bertone}, {Bianchi}, {Binnenfeld}, {Blanco-Cuaresma}, {Blazere}, {Boch}, {Bombrun}, {Bossini}, {Bouquillon}, {Bragaglia}, {Bramante}, {Breedt},
  {Bressan}, {Brouillet}, {Brugaletta}, {Bucciarelli}, {Burlacu}, {Butkevich}, {Buzzi}, {Caffau}, {Cancelliere}, {Cantat-Gaudin}, {Carballo}, {Carlucci}, {Carnerero}, {Carrasco}, {Casamiquela}, {Castellani}, {Castro-Ginard}, {Chaoul}, {Charlot}, {Chemin}, {Chiaramida}, {Chiavassa}, {Chornay}, {Comoretto}, {Contursi}, {Cooper}, {Cornez}, {Cowell}, {Crifo}, {Cropper}, {Crosta}, {Crowley}, {Dafonte}, {Dapergolas}, {David}, {David}, {de Laverny}, {De Luise}, {De March}, {De Ridder}, {de Souza}, {de Torres}, {del Peloso}, {del Pozo}, {Delbo}, {Delgado}, {Delisle}, {Demouchy}, {Dharmawardena}, {Di Matteo}, {Diakite}, {Diener}, {Distefano}, {Dolding}, {Edvardsson}, {Enke}, {Fabre}, {Fabrizio}, {Faigler}, {Fedorets}, {Fernique}, {Fienga}, {Figueras}, {Fournier}, {Fouron}, {Fragkoudi}, {Gai}, {Garcia-Gutierrez}, {Garcia-Reinaldos}, {Garc{\'\i}a-Torres}, {Garofalo}, {Gavel}, {Gavras}, {Gerlach}, {Geyer}, {Giacobbe}, {Gilmore}, {Girona}, {Giuffrida}, {Gomel}, {Gomez}, {Gonz{\'a}lez-N{\'u}{\~n}ez},
  {Gonz{\'a}lez-Santamar{\'\i}a}, {Gonz{\'a}lez-Vidal}, {Granvik}, {Guillout}, {Guiraud}, {Guti{\'e}rrez-S{\'a}nchez}, {Guy}, {Hatzidimitriou}, {Hauser}, {Haywood}, {Helmer}, {Helmi}, {Sarmiento}, {Hidalgo}, {Hilger}, {H{\l}adczuk}, {Hobbs}, {Holland}, {Huckle}, {Jardine}, {Jasniewicz}, {Jean-Antoine Piccolo}, {Jim{\'e}nez-Arranz}, {Jorissen}, {Juaristi Campillo}, {Julbe}, {Karbevska}, {Kervella}, {Khanna}, {Kontizas}, {Kordopatis}, {Korn}, {K{\'o}sp{\'a}l}, {Kostrzewa-Rutkowska}, {Kruszy{\'n}ska}, {Kun}, {Laizeau}, {Lambert}, {Lanza}, {Lasne}, {Le Campion}, {Lebreton}, {Lebzelter}, {Leccia}, {Leclerc}, {Lecoeur-Taibi}, {Liao}, {Licata}, {Lindstr{\o}m}, {Lister}, {Livanou}, {Lobel}, {Lorca}, {Loup}, {Madrero Pardo}, {Magdaleno Romeo}, {Managau}, {Mann}, {Manteiga}, {Marchant}, {Marconi}, {Marcos}, {Marcos Santos}, {Mar{\'\i}n Pina}, {Marinoni}, {Marocco}, {Marshall}, {Martin Polo}, {Mart{\'\i}n-Fleitas}, {Marton}, {Mary}, {Masip}, {Massari}, {Mastrobuono-Battisti}, {Mazeh}, {McMillan}, {Messina}, {Michalik},
  {Millar}, {Mints}, {Molina}, {Molinaro}, {Moln{\'a}r}, {Monari}, {Mongui{\'o}}, {Montegriffo}, {Montero}, {Mor}, {Mora}, {Morbidelli}, {Morel}, {Morris}, {Muraveva}, {Murphy}, {Musella}, {Nagy}, {Noval}, {Oca{\~n}a}, {Ogden}, {Ordenovic}, {Osinde}, {Pagani}, {Pagano}, {Palaversa}, {Palicio}, {Pallas-Quintela}, {Panahi}, {Payne-Wardenaar}, {Pe{\~n}alosa Esteller}, {Penttil{\"a}}, {Pichon}, {Piersimoni}, {Pineau}, {Plachy}, {Plum}, {Poggio}, {Pr{\v{s}}a}, {Pulone}, {Racero}, {Ragaini}, {Rainer}, {Raiteri}, {Rambaux}, {Ramos}, {Ramos-Lerate}, {Re Fiorentin}, {Regibo}, {Richards}, {Rios Diaz}, {Ripepi}, {Riva}, {Rix}, {Rixon}, {Robichon}, {Robin}, {Robin}, {Roelens}, {Rogues}, {Rohrbasser}, {Romero-G{\'o}mez}, {Rowell}, {Royer}, {Ruz Mieres}, {Rybicki}, {Sadowski}, {S{\'a}ez N{\'u}{\~n}ez}, {Sagrist{\`a} Sell{\'e}s}, {Sahlmann}, {Salguero}, {Samaras}, {Sanchez Gimenez}, {Sanna}, {Santove{\~n}a}, {Sarasso}, {Schultheis}, {Sciacca}, {Segol}, {Segovia}, {S{\'e}gransan}, {Semeux}, {Shahaf}, {Siddiqui}, {Siebert},
  {Siltala}, {Silvelo}, {Slezak}, {Slezak}, {Smart}, {Snaith}, {Solano}, {Solitro}, {Souami}, {Souchay}, {Spagna}, {Spina}, {Spoto}, {Steele}, {Steidelm{\"u}ller}, {Stephenson}, {S{\"u}veges}, {Surdej}, {Szabados}, {Szegedi-Elek}, {Taris}, {Taylor}, {Teixeira}, {Tolomei}, {Tonello}, {Torra}, {Torra}, {Torralba Elipe}, {Trabucchi}, {Tsounis}, {Turon}, {Ulla}, {Unger}, {Vaillant}, {van Dillen}, {van Reeven}, {Vanel}, {Vecchiato}, {Viala}, {Vicente}, {Voutsinas}, {Weiler}, {Wevers}, {Wyrzykowski}, {Yoldas}, {Yvard}, {Zhao}, {Zorec}, {Zucker}, \& {Zwitter}}]{2023AA...674A...1G}
{Gaia Collaboration}, {Vallenari}, A., {Brown}, A.~G.~A., {et~al.} 2023, {Gaia Data Release 3. Summary of the content and survey properties}, \aap, 674, A1, \dodoi{10.1051/0004-6361/202243940}

\bibitem[{{Gordon} {et~al.}(2023){Gordon}, {Clayton}, {Decleir}, {Fitzpatrick}, {Massa}, {Misselt}, \& {Tollerud}}]{gordon2023}
{Gordon}, K.~D., {Clayton}, G.~C., {Decleir}, M., {et~al.} 2023, {One Relation for All Wavelengths: The Far-ultraviolet to Mid-infrared Milky Way Spectroscopic R(V)-dependent Dust Extinction Relationship}, \apj, 950, 86, \dodoi{10.3847/1538-4357/accb59}

\bibitem[{{Mack} {et~al.}(2021){Mack}, {Olszewksi}, \& {Pirzkal}}]{mack2021}
{Mack}, J., {Olszewksi}, H., \& {Pirzkal}, N. 2021, {WFC3/IR Filter-Dependent Sky Flats}, WFC3 Instrument Science Report 2021-01, 29 pages

\bibitem[{{Marinelli} {et~al.}(2024){Marinelli}, {Bajaj}, {Calamida}, \& {Mack}}]{2024wfc..rept....6M}
{Marinelli}, M., {Bajaj}, V., {Calamida}, A., \& {Mack}, J. 2024, {Time-Dependent Sensitivity of the WFC3/IR Detector}, WFC3 Instrument Science Report 2024-06, 56 pages

\bibitem[{{M{\'e}sz{\'a}ros} {et~al.}(2024){M{\'e}sz{\'a}ros}, {Bohlin}, {Allende Prieto}, {Cseh}, {Kov{\'a}cs}, {Fleming}, {Dencs}, {Deustua}, {Gordon}, {Hubeny}, {Mez{\H{o}}}, \& {Truszek}}]{meszaros2024}
{M{\'e}sz{\'a}ros}, S., {Bohlin}, R., {Allende Prieto}, C., {et~al.} 2024, {The updated BOSZ synthetic stellar spectral library}, arXiv e-prints, arXiv:2407.10872, \dodoi{10.48550/arXiv.2407.10872}

\bibitem[{{Mullally} {et~al.}(2022){Mullally}, {Sloan}, {Hermes}, {Kunz}, {Hambleton}, {Bohlin}, {Fleming}, {Gordon}, {Kaleida}, \& {Mohamed}}]{mullally2022}
{Mullally}, S.~E., {Sloan}, G.~C., {Hermes}, J.~J., {et~al.} 2022, {Searching for TESS Photometric Variability of Possible JWST Spectrophotometric Standard Stars}, \aj, 163, 136, \dodoi{10.3847/1538-3881/ac4bce}

\bibitem[{{Som} {et~al.}(2024){Som}, {Bohlin}, {Mack}, {Bajaj}, \& {Calamida}}]{som2024}
{Som}, D., {Bohlin}, R., {Mack}, J., {Bajaj}, V., \& {Calamida}, A. 2024, {Sensitivity Evolution of WFC3/IR Using Spatial Scanning Photometry and Grism Spectrophotometry}, WFC3 Instrument Science Report 2024-01, 12 pages

\bibitem[{{STScI Development Team}(2018)}]{2018ascl.soft11001S}
{STScI Development Team}. 2018, {synphot: Synthetic photometry using Astropy}, Astrophysics Source Code Library, record ascl:1811.001.
\newblock \doeprint{1811.001}

\bibitem[{{STScI Development Team}(2020)}]{2020ascl.soft10003S}
---. 2020, {stsynphot: synphot for HST and JWST}, Astrophysics Source Code Library, record ascl:2010.003.
\newblock \doeprint{2010.003}

\end{thebibliography}
\bibliographystyle{aasjournal}
\nocite{*}

\end{document}